\documentclass[journal,onecolumn]{IEEEtran}
\UseRawInputEncoding
\usepackage[ruled,linesnumbered]{algorithm2e}

\usepackage{color}
\usepackage{makecell}

\usepackage{subfigure,array}
\usepackage{graphicx,cite,epsfig,amssymb,amsmath,multirow,lettrine,flushend,extarrows}

\begin{document}

\title{Optimal Operation of a Hydrogen-based Building Multi-Energy System Based on Deep Reinforcement Learning}
\author{{Liang~Yu,~\IEEEmembership{Member,~IEEE}, Shuqi Qin, Zhanbo Xu,~\IEEEmembership{Member,~IEEE}, Xiaohong~Guan,~\IEEEmembership{Fellow,~IEEE},\\ Chao Shen,~\IEEEmembership{Senior Member,~IEEE}, and Dong~Yue,~\IEEEmembership{Fellow,~IEEE}}
\thanks{\newline L. Yu is with Xi'an Jiaotong University, Xi'an 710049, China, and is also with Nanjing University of Posts and Telecommunications, Nanjing 210003, China. (email: liang.yu@njupt.edu.cn) \newline
Z. Xu, X. Guan, and C. Shen are with Systems Engineering Institute, Ministry of Education Key Lab for Intelligent Networks and Network Security, Xi'an Jiaotong University, Xi'an 710049, China. (email: \{zbxu,xhguan,cshen\}@sei.xjtu.edu.cn) \newline
S. Qin and D. Yue are with Nanjing University of Posts and Telecommunications, Nanjing 210003, China. (email: shuqi\_qin@163.com, medongy@vip.163.com)\newline
}}


\maketitle

\begin{abstract}
Since hydrogen has many advantages (e.g., free pollution, extensive sources, convenient storage and transportation), hydrogen-based multi-energy systems (HMESs) have received wide attention. However, existing works on the optimal operation of HMESs neglect building thermal dynamics, which means that the flexibility of building thermal loads can not be utilized for reducing system operation cost. In this paper, we investigate an optimal operation problem of an HMES with the consideration of building thermal dynamics. Specifically, we first formulate an expected operational cost minimization problem related to an HMES. Due to the existence of uncertain parameters, inexplicit building thermal dynamics models, temporally coupled operational constraints related to three kinds of energy storage systems and indoor temperatures, as well as the coupling between electric energy subsystems and thermal energy subsystems, it is challenging to solve the formulated problem. To overcome the challenge, we reformulate the problem as a Markov game and propose an energy management algorithm to solve it based on multi-agent discrete actor-critic with rules (MADACR). Note that the proposed algorithm does not require any prior knowledge of uncertain parameters, parameter prediction, and explicit building thermal dynamics model. Simulation results based on real-world traces show the effectiveness of the proposed algorithm.
\end{abstract}


\section{Introduction}\label{s1}
Buildings account for a large portion of total energy consumption and total carbon emission in the world. For example, global buildings consumed about 30\% of the total energy and generated about 28\% of the total carbon emission in 2019\cite{Global2020}. Since the global energy supply mainly depends on fossil fuels, energy and environmental issues are incurred\cite{Liu2021}. Due to many advantages (e.g., free pollution, extensive sources, convenient storage and transportation), hydrogen energy has attracted widespread attention and is recognized as a promising alternative to fossil fuels\cite{Liu2021,Pan2020,Ozturk2020}. Moreover, the coordination of hydrogen energy storage system (HESS) and other energy storage systems (ESSs) (e.g., thermal energy storage and electric energy storage) contributes to the improvement of building energy efficiency\cite{Liu2021}. Therefore, it is of great importance to optimize the operation of a hydrogen-based building multi-energy system (HBMES).

In the literature, many approaches have been proposed for the planning or operation of multi-energy systems, e.g., mixed-integer linear programming (MILP)\cite{Huang2019}, nonconvex quadratically constrained programming\cite{Lu2020}, stochastic programming\cite{Sharma2019}\cite{Li2020}, robust optimization\cite{Sharma2021}\cite{Cesena2019}\cite{Lu2021}, Benders decomposition\cite{Zheng2021}, model-predictive control (MPC)\cite{Ceusters2021}\cite{Jin2021}, and deep reinforcement learning\cite{Ye2020}. For example, Lu \emph{et al.} proposed an algorithm to minimize system operation cost of heat and electricity integrated energy systems based on two-stage robust programming\cite{Lu2021}. In \cite{Ye2020}, Ye \emph{et al.} developed a real-time energy management approach to minimize the energy cost of a residential multi-energy system based on prioritized deep deterministic policy gradient. Although some efforts have been made, the above-mentioned studies do not consider the utilization of hydrogen energy storage. To promote the development of hydrogen energy storage, some works have investigated the optimal planning or operation problem of hydrogen-based multi-energy systems\cite{Liu2021}\cite{Pan2020}\cite{Mehrjerdi2020,Langeroudi2021,Dong2020,Tao2020} and proposed many optimization approaches, e.g., MILP\cite{Liu2021}, two-stage stochastic programming\cite{Langeroudi2021}, mixed integer programming\cite{Dong2020}, two-stage robust optimization\cite{Pan2020}, and distributed optimization\cite{Tao2020}. In existing works on the optimal operation of hydrogen-based multi-energy systems, building thermal dynamics and thermal comfort of occupants are neglected, which means that the flexibility of building thermal loads can not be utilized for reducing system operation cost. Moreover, the approaches proposed in above-mentioned studies neglect the system uncertainty\cite{Mehrjerdi2020}\cite{Dong2020} or require prior knowledge of uncertain parameters\cite{Langeroudi2021}, or need to forecast uncertain parameters\cite{Pan2020}.

Based on the above observation, we investigate an optimal operation problem related to an HBMES with the consideration of building thermal dynamics and intend to propose an efficient energy management approach, which can cope with the limitations of existing approaches mentioned above. To be specific, we intend to minimize the expected operational cost of an HBMES by intelligently scheduling thermal loads and various ESSs, including hydrogen, thermal, and electric ESSs. However, several challenges are involved in achieving the above aim. Firstly, there are many uncertain parameters, e.g., renewable generation output, electric load, electricity price, outdoor temperature, and carbon emission rate. Secondly, there are some temporally coupled operational constraints related to several ESSs and indoor temperatures. Thirdly, there is coupling between electricity and heat introduced by the HESS, which means that HESS has to coordinate with two kinds of energy subsystems for minimizing operational cost. Finally, it is difficult to obtain explicit building thermal dynamics models that are accurate and efficient enough for building control. Even if such models exist, it is often a time-consuming and error-prone process to develop and maintain them\cite{Wei2017,Gao2020,YuJIOT2020}. Taking the above challenges into consideration, existing methods that need to know explicit building thermal dynamics models are not applicable, e.g., model predictive control\cite{Ma2015}, Lyapunov optimization techniques\cite{LiangTSG2019}, and robust optimization\cite{Sharma2021}. To overcome these challenges, we propose a real-time energy management algorithm based on multi-agent deep reinforcement learning (MADRL)\cite{NguyenMAS2019}, which is helpful for the efficient collaboration among all energy subsystems under uncertainties\cite{YuJIOT2021}.

The main contributions of this paper are summarized as follows.
\begin{itemize}
  \item We formulate an expected operational cost minimization problem for an HBMES with the consideration of three kinds of ESSs and building thermal dynamics. Due to the existence of above-mentioned challenges, we reformulate the cost minimization problem as a Markov game.
  \item We propose a real-time energy management algorithm to solve the Markov game based on multi-agent discrete actor-critic with rules (MADACR). Note that the proposed algorithm does not require any prior knowledge of uncertain parameters, parameter prediction, and explicit building thermal dynamics models.
  \item Simulation results based on real-world traces show that the proposed algorithm can reduce operational cost by 8.1\%-44.26\% compared with other baselines while still maintaining comfortable temperature ranges.
\end{itemize}

The rest of this paper is organized as follows. In Section~\ref{s2}, we introduce the system model and formulate an expected operational cost minimization problem. In Section~\ref{s3}, we reformulate the minimization problem as a Markov game. In Section~\ref{s4}, we propose an energy management algorithm to solve the Markov game. In Section~\ref{s5}, performance evaluations are conducted. Finally, we draw a conclusion and point out the future work in Section~\ref{s6}.

\section{System Model And Problem Formulation}\label{s2}
\begin{figure}[!htb]
\centering
\includegraphics[scale=0.52]{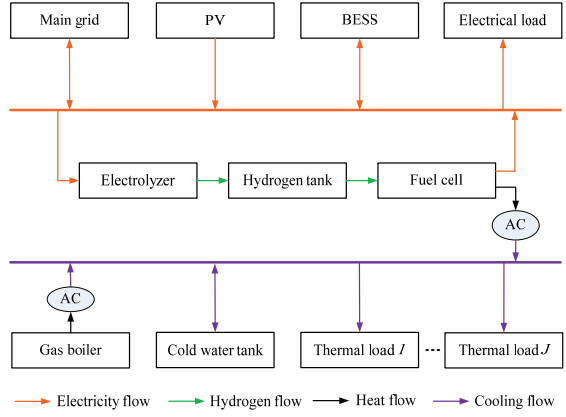}
\caption{Illustration of an HBMES}\label{fig_1}
\end{figure}

We consider an HBMES in Fig.~\ref{fig_1}, where main grid, photovoltaic (PV) generation, battery energy storage system (BESS), electrical load, electrolyzer, hydrogen tank, fuel cell, gas boiler, cold water tank (CWT), and thermal loads can be identified. Among these components, there are four kinds of energy flows, i.e., electricity flow, hydrogen flow, heat flow, and cooling flow. In electricity flow, electrical load (e.g., electric vehicles, electric water heaters, and computers) can be served by main grid, PV generators, BESS, and fuel cell. Moreover, it can be seen that hydrogen flow appears in HESS, which consists of electrolyzer, hydrogen tank, and fuel cell. To be specific, the hydrogen generated by electrolyzer can be stored in the hydrogen tank, which will discharge hydrogen to drive fuel cell for generating electricity and heat simultaneously. The heat generated by fuel cell and gas boiler can be transformed into cold water by absorption chiller (AC). Next, cold water can be stored in CWT and used for cooling buildings. In the following parts, we first introduce the models related to PV generation, gas boiler, energy storage, thermal load, power/energy balance, and operational cost. Then, we formulate an expected operational cost minimization problem related to HBMES.

\subsection{PV Generation Model}\label{s21}
Let $P_{\text{pv},t}$ be the maximum generation output of PV system at slot $t$. Then, its value can be estimated by\cite{LiangTSG2019}
\begin{equation}\label{f_1}
P_{\text{pv},t}=\eta_{\text{pv}}{h_{\text{pv}}}{\kappa_{l,t}},
\end{equation}
where $\eta_{\text{pv}}$ denotes the PV system generation efficiency, $h_{\text{pv}}$ is total radiation area of solar panels, and ${\kappa_{l,t}}$ is the solar radiation intensity at slot $t$.


\subsection{Gas Boiler Model}\label{s22}
Let $P_{\text{gb},t}$ denotes the heat power output of the gas boiler at slot $t$. Then, we have
\begin{equation}\label{f_2}
0\leq P_{\text{gb},t} \leq P^{\max}_{\text{gb}},
\end{equation}
where $P^{\max}_{\text{gb}}$ is the maximum heat power output of gas boiler.

\subsection{Energy Storage Model}\label{s23}

\subsubsection{Battery Energy Storage Model}
Let $B_t$ be the stored energy level in the BESS at slot $t$. Then, the dynamics of energy level in BESS can be described by\cite{YuJIOT2020}
\begin{equation}\label{f_3}
B_{t+1}=B_t + (\eta_{\text{bc}} P_{\text{bc},t}+\frac{P_{\text{bd},t}}{\eta_{\text{bd}}})\Delta t,
\end{equation}
where $\eta_{\text{bc}}$ and $\eta_{\text{bd}}$ are the charging and discharging efficiency coefficients, respectively; $P_{\text{bc},t}$ and $P_{\text{bd},t}$ are charging power and discharging power of BESS, respectively. Here, $P_{\text{bc},t}$ and $P_{\text{bd},t}$ are assigned with different signs (i.e., $P_{\text{bc},t}\geq 0$ and $P_{\text{bd},t} \leq 0$), which are helpful for the action design in Section~\ref{s3}.

To ensure that the energy level of the BESS fluctuates within a normal range at any time, we have
\begin{equation}\label{f_4}
B^{\min} \le B_{t} \le B^{\max},
\end{equation}
where $B^{\min}$ and $B^{\max}$ be the minimum and maximum energy levels of the BESS, respectively.

Let $P_{\text{bc}}^{\max}$ and $P_{\text{bd}}^{\max}$ be the maximum charging power and maximum discharging power, respectively. Then, we have
\begin{equation}\label{f_5}
0 \le P_{\text{bc},t} \le P_{\text{bc}}^{\max},
\end{equation}
\begin{equation}\label{f_6}
-P_{\text{bd}}^{\max} \le P_{\text{bd},t} \le 0.
\end{equation}

Taking the round-trip inefficiency into consideration, simultaneous charging and discharging is not allowed, which can be depicted by
\begin{equation}\label{f_7}
P_{\text{bc},t} \cdot {P_{\text{bd},t}} = 0.
\end{equation}

\subsubsection{Thermal Energy Storage Model}

Let $Q_{\text{th},t}$ be the stored thermal energy in CWT at slot $t$. Then, its dynamics can be described by
\begin{equation}\label{f_8}
Q_{\text{th},t+1}=Q_{\text{th},t}+(P_{\text{tc},t}\eta_{\text{tc}}+\frac{P_{\text{td},t}}{\eta_{\text{td}}})\Delta t,
\end{equation}
where $\eta_{\text{tc}}$ and $\eta_{\text{td}}$ are injection efficiency and release efficiency of CWT, respectively; $P_{\text{tc},t}$ and $P_{\text{td},t}$ are injected power and released power at slot $t$, respectively. Here, $P_{\text{tc},t}$ and $P_{\text{td},t}$ are assigned with different signs (i.e., $P_{\text{tc},t}\geq 0$ and $P_{\text{td},t} \leq 0$), which are helpful for the action design in Section~\ref{s3}.

To ensure the normal operation of CWT, the following operational constraints of the CWT should be satisfied, i.e.,
\begin{equation}\label{f_9}
0\leq Q_{\text{th},t}\leq Q_{\text{th}}^{\max},
\end{equation}
\begin{equation}\label{f_10}
-P_{\text{td}}^{\max} \leq P_{\text{td},t}\leq 0,
\end{equation}
\begin{equation}\label{f_11}
0\leq P_{\text{tc},t}\leq P_{\text{tc}}^{\max},
\end{equation}
\begin{equation}\label{f_12}
P_{\text{td},t}\cdot P_{\text{tc},t}=0,
\end{equation}
where $Q_{\text{th}}^{\max}$ denotes the capacity of the CWT; $P_{\text{td}}^{\max}$ and $P_{\text{tc}}^{\max}$ are the maximum released power and injected power, respectively. \eqref{f_9} denotes that the stored thermal energy level should fluctuate within a feasible range. \eqref{f_10} and \eqref{f_11} denotes the effective range of released power and injected power, respectively. \eqref{f_12} means that releasing and injecting cold water can not happen simultaneously so that meaningless thermal loss can be avoided.

\subsubsection{Hydrogen Energy Storage Model}

Let $H_{t}$ be the storage level of hydrogen in the tank (in \text{Nm$^3$}). Then, the dynamics of hydrogen storage level can be described by
\begin{equation}\label{f_13}
H_{t+1} = H_t +P_{\text{el},t} \omega_{\text{el}}\Delta t+\frac{P_{\text{fc},t}}{\omega_{\text{fc}}}\Delta t,
\end{equation}
where $P_{\text{el},t}$ and $P_{\text{fc},t}$ are charging power of the electrolyzer and discharging power of fuel cell, respectively; ${\omega_{\text{el}}}$ (in \text{Nm$^3$/kWh}) and ${\omega_{\text{fc}}}$ (in \text{kWh/Nm$^3$}) denote the conversion coefficients of electrolyzer and fuel cell, respectively; Here, $P_{\text{el},t}$ and $P_{\text{fc},t}$ are assigned with different signs (i.e., $P_{\text{el},t}\geq 0$ and $P_{\text{fc},t} \leq 0$), which are helpful for the action design in Section~\ref{s3}.

Since the maximum storage level of the hydrogen tank is limited by its tolerable tank pressure\cite{Liu2021}\cite{Wu2019}, we have
\begin{equation}\label{f_14}
0\leq H_t \le H^{\max},
\end{equation}
where $H^{\max}$ is the storage capacity of the hydrogen tank.

To keep the efficiency of the HESS, we assume that electrolyzer and fuel cell can not operate simultaneously. Then, we have
\begin{equation}\label{f_15}
P_{\text{el},t} \cdot P_{\text{fc},t}=0.
\end{equation}

In addition, the power consumption of the electrolyzer and electric power output of fuel cell should satisfy the following physical constraints, which can be given by
\begin{equation}\label{f_16}
0 \le P_{\text{el},t } \le P^{\max}_{\text{el}},
\end{equation}
\begin{equation}\label{f_17}
-P^{\max}_{\text{fc}} \le P_{\text{fc},t } \le 0,
\end{equation}
where $P^{\max}_{\text{el}}$ and $P^{\max}_{\text{fc}}$ are the rated powers of electrolyzer and fuel cell, respectively.

Since the fuel cell generates electricity and heat simultaneously, the electrical output power of the fuel cell $P_{\text{fc},t}$ is coupled with the corresponding thermal output power $Q_{\text{fc},t}$. Then, we have\cite{Liu2021}
\begin{equation}\label{f_18}
Q_{\text{fc},t}=\eta_{\text{hr}}\eta_{\text{h2e}}P_{\text{fc},t}\Delta t,
\end{equation}
where $\eta_{\text{h2e}}$ and $\eta_{\text{hr}}$ are heat-to-electricity ratio and the heat recovery efficiency, respectively.

\subsection{Thermal Load Model}\label{s24}
Let $P_{\text{sp},i,t}$ be the thermal input power for cooling demand in building $i$ at slot $t$, which will affect building indoor temperature $\beta_{\text{in},i,t}$. To provide a comfortable temperature range for occupants in building $i$, the following constraints should be satisfied,
\begin{equation}\label{f_19}
\beta_i^{\min}\leq \beta_{\text{in},i,t} \leq \beta_i^{\max},
\end{equation}
\begin{equation}\label{f_20}
\beta_{\text{in},i,t+1}=\mathcal{F}_i(P_{\text{sp},i,t},~\beta_{\text{out},t},~\beta_{\text{in},i,t},~\varrho_{i,t}),
\end{equation}
\begin{equation}\label{f_21}
0\leq P_{\text{sp},i,t}\leq P_{\text{sp},i}^{\max},
\end{equation}
where $\beta_i^{\min}$ and $\beta_i^{\max}$ are the lower limit and upper limit of the comfortable temperature rage in building $i$, respectively; $\beta_{\text{out},t}$ and $\varrho_{i,t}$ are outdoor temperature and random thermal disturbance at slot $t$, respectively; $\mathcal{F}_i(\cdot)$ denotes a thermal dynamics model of building $i$, and $P_{\text{sp},i}^{\max}$ denotes the maximum thermal input power in building $i$.

\subsection{Power/Energy Balance Model}\label{s25}
To maintain the electric power balance at each slot $t$, we have
\begin{equation}\label{f_22}
P_{\text{g},t} + P_{\text{pv},t} - P_{\text{fc},t} - P_{\text{bd},t}=P_{\text{el},t} + P_{\text{load},t} + P_{\text{bc},t},
\end{equation}
where $P_{\text{g},t}$ represents the power interaction between HBMES and the main grid at slot $t$; $P_{\text{load},t}$ denotes the power demand at slot $t$. To be specific, $P_{\text{g},t}>0$ means that buying electricity from the main grid at slot $t$. Otherwise, electricity will be sold to the main grid at slot $t$.

Similarly, thermal energy balance at each slot $t$ can be depicted by the following constraint, i.e.,
\begin{equation}\label{f_23}
Q_{\text{fc},t}\eta_{\text{h2c}}\geq (P_{\text{tc},t}+P_{\text{td},t}+\sum_{i=1}^{N_b}P_{\text{sp},i,t}-P_{\text{gb},t}\eta_{\text{h2c}})\Delta t,
\end{equation}
where $N_b$ denotes the number of buildings and $\eta_{\text{h2c}}$ denotes AC transformation efficiency from heating to cooling.

\subsection{Operational Cost Model}\label{s26}
The operational cost of the HBMES consists of six parts, i.e., the energy cost of electricity buying or selling $C_{1,t}$, carbon emission cost $C_{2,t}$, BESS depreciation cost $C_{3,t}$, HESS related cost $C_{4,t}$, CWT depreciation cost $C_{5,t}$, and gas purchasing cost $C_{6,t}$.

Let ${v_t}$ and $\tau_t$ be the buying and selling prices of electricity, respectively. Then, $C_{1,t}$ is expressed by
\begin{equation}\label{f_24}
C_{1,t}=  \Big(\frac{v_t-\tau_t}{2}\left|P_{\text{g},t}\right| + \frac{v_t + \tau_t}{2}P_{\text{g},t}\Big)\Delta t,
\end{equation}
where $C_{1,t} = v_{t}P_{\text{g},t}$ if $P_{\text{g},t} \geq 0$; Otherwise, $C_{1,t}=\tau_{t}P_{\text{g},t}$.

Let $\mu_{\text{e},t}$ (in kg/kWh) be the carbon emission rate of the main grid at slot $t$. Then, the carbon emission generated by the HBMES at slot $t$ can be given by $\mu_{\text{e},t}P_{\text{g},t}\Delta t$. Then, the carbon emission cost is calculated by\cite{Dong2020}
\begin{equation}\label{f_25}
C_{2,t}=\mu_{\text{c}}\mu_{\text{e},t}P_{\text{g},t}\Delta t,
\end{equation}
where $\mu_{\text{c}}$ is a weighted parameter in RMB/kg, which denotes the importance of carbon emission with respect to energy cost.

Since too frequent charging or discharging will damage the life of the BESS, BESS depreciation cost is adopted\cite{YuJIOT2020}, i.e.,
\begin{equation}\label{f_26}
C_{3,t} = \psi_{\text{BESS}} (\left| {{P_{\text{bc},t}}} \right| + \left| {{P_{\text{bd},t}}} \right|),
\end{equation}
where $\psi_{\text{BESS}}$ is the battery depreciation coefficient in RMB/kW.

According to \cite{Garcia2015}, the startup and shutdown cycles have degradation effects on electrolyzer and fuel cell. Thus, startup and shutdown costs are considered in this paper. Let $\delta_{\text{x}}^{\text{on}}$, $\delta_{\text{x}}^{\text{su}}$, and $\delta_{\text{x}}^{\text{sd}}$ be the operation cost, startup cost, and shutdown cost of component $\text{x}$ ($\text{x}\in\{\text{el},\text{fc}\}$) in HESS, respectively, where ``$\text{el}$" and ``$\text{fc}$" denote electrolyzer and fuel cell, respectively. Then, $C_{4,t}$ can be calculated by\cite{Garcia2015}
\begin{align}\label{f_27}
C_{4,t} =\sum\nolimits_{\text{x}\in\{\text{el},\text{fc}\}}\delta_{\text{x}}^{\text{on}}I_{\text{x},t}^{\text{on}}+\delta_{\text{x}}^{\text{su}}I_{\text{x},t}^{\text{su}}+\delta_{\text{x}}^{\text{sd}}I_{\text{x},t}^{\text{sd}},
\end{align}
where $I_{\text{x},t}^{\text{on}}$, $I_{\text{x},t}^{\text{su}}$, and $I_{\text{x},t}^{\text{sd}}$ are logical indicator variables related to ON/OFF state, startup state, and shutdown state of component $\text{x}$, respectively; $I_{\text{x},t}^{\text{su}}=\max\{I_{\text{x},t}^{\text{on}}-I_{\text{x},t-1}^{\text{on}},0\}$ and $I_{\text{x},t}^{\text{sd}}=\max\{I_{\text{x},t-1}^{\text{on}}-I_{\text{x},t}^{\text{on}},0\}$.

Similar to BESS, CWT depreciation cost can be captured by\cite{YuJIOT2020}
\begin{equation}\label{f_28}
C_{5,t} = \psi_{\text{CWT}} (\left| {{P_{\text{tc},t}}} \right| + \left| {{P_{\text{td},t}}} \right|),
\end{equation}
where $\psi_{\text{CWT}}$ is the CWT depreciation coefficient in RMB/kW.

Let $\eta_{\text{gb}}$ and $\lambda_{g,t}$ be gas-to-heat conversion efficiency and gas price (in $\text{RMB}/\text{kWh}$), respectively. Then, the gas purchasing cost at slot $t$ can be given by\cite{Ye2020}
\begin{equation}\label{f_29}
C_{6,t}=\lambda_{g,t}\frac{P_{\text{gb},t}\Delta t}{\eta_{\text{gb}}}.
\end{equation}

\subsection{Expected Operational Cost Minimization Problem}\label{s27}
Based on above models, we can formulate an expected operational cost minimization problem of an HBMES as follows,
\begin{subequations}\label{f_30}
\begin{align}
(\textbf{P1})~&\min\sum\limits_{t = 0}^{T-1}\mathbb{E}\Big\{\sum\limits_{j=1}^6 C_{j,t}\Big\}\\
s.t.&~\eqref{f_1}-\eqref{f_29},
\end{align}
\end{subequations}
where the expectation operator $\mathbb{E}$ is taken over the randomness of system parameters (i.e., PV generation output $P_{\text{pv},t}$, power demand $P_{\text{load},t}$, carbon emission rate $\mu_{\text{e},t}$, thermal load $Q_{\text{load},t}$, buying/selling price $v_{t}/\tau_{t}$), and possible stochastic control decisions (i.e., $P_{\text{gb},t}$, $P_{\text{bc},t}$, $P_{\text{bd},t}$, $P_{\text{tc},t}$, $P_{\text{td},t}$, $P_{\text{el},t}$, $P_{\text{fc},t}$, $P_{\text{sp},i,t}|_{1\leq i \leq J}$, and $P_{\text{g},t}$).

Solving \textbf{P1} is a nontrivial task due to the following reasons. Firstly, there are many uncertain parameters and it is often difficult to know their statistical distributions of all combinations in practice. Secondly, there are several temporally coupled operational constraints (e.g., \eqref{f_3}, \eqref{f_8}, \eqref{f_13}, and \eqref{f_20}). Thirdly, there is a coupling between electricity and heat incurred by the HESS, which means that HESS has to coordinate with electric--flow subsystems and thermal-energy-flow subsystems for jointly minimizing the operational cost. Finally, it is challenging to obtain an explicit building thermal dynamics model $\mathcal{F}_i(\cdot)$ that is accurate and efficient enough for building control\cite{Wei2017}.

To address the first challenge, some methods can be adopted, e.g., stochastic programming, robust optimization, Lyapunov optimization techniques, and model predictive control. However, these methods either need to know prior knowledge (e.g., probability distribution, maximum and minimum values) of uncertain parameters or predict/approximate random parameters. To deal with the second challenge, typical methods are based on dynamic programming\cite{YuJIOT2020}, which suffers from ``the curse of dimensionality" problem. To overcome the last challenge, many model-free DRL (note that DRL has powerful ability of deep learning\cite{LiZ2021} and strong decision-making ability of reinforcement learning\cite{LiYuanzheng2021}) methods can be adopted\cite{Wei2017}\cite{Gao2020}\cite{YuJIOT2020}, which can help agents to learn optimal policies from the process of interacting with building environments. Once the optimal policies are found, they can operate without knowing any prior information of uncertain parameters, parameter prediction, and explicit building thermal dynamics models. However, these DRL methods can not be applied to \textbf{P1} directly due to the coupling between electricity and heat incurred by HESS, which can be seen in \eqref{f_18}, \eqref{f_22}, and \eqref{f_23}. Based on the above analysis, we are motivated to design an energy management algorithm based on MADRL, which can support scalable cooperation among different energy subsystems under uncertainties by designing proper reward function and appropriate algorithms\cite{YuJIOT2021}. Thus, we intend to reformulate \textbf{P1} as a Markov game in Section~\ref{s3} and design an MADRL-based algorithm in Section~\ref{s4}.

\section{Problem Reformulation}\label{s3}
In this section, we reformulate \textbf{P1} as a Markov game, which is a general modeling framework for multi-agent decision-making problem under uncertainty\cite{NguyenMAS2019}. Specifically, a Markov game can be defined by a set of states, $S$, a collection of action sets (each action set is associated with each agent in the environment), $A_1$,~$\cdots$,~$A_{N}$, a state transition function, $F:~S\times A_1\times \ldots \times A_{N}\rightarrow \Pi(S)$, which defines the probability distribution over possible next states, given the current state and actions for all agents, and a reward function for each agent $i$ ($1\leq i \leq N$), $R_i:~S\times A_1\times \ldots \times A_{N}\rightarrow \mathbb{R}$. In a Markov game, each agent $i$ takes action $a_i\in A_i$ based on its local observation $o_i\in \mathcal{O}_i$, where $o_i$ contains partial information of the global state $s\in S$. The aim of the agent $i$ is to maximize its expected return by learning a policy $\pi_i:~\mathcal{O}_i\rightarrow \Pi(A_i)$, which maps the agent's local observation $o_i\in \mathcal{O}_i$ into a distribution over its set of actions. Here, the return is the sum of discounted rewards received over the future, i.e., $\sum\nolimits_{j=0}^{\infty}\gamma^j r_{i,t+j+1}(s_t,a_{1,t},\cdots,a_{N,t})$, where $\gamma\in[0,1]$ is a discount factor and $r_{i,t+1}\in R_i$ is the reward received by the agent $i$ at slot $t$. Since the information of state transition function is not required in the proposed algorithm, we design three components of the Markov game, i.e., state, action, and reward function.

Typically, the number of agents is the same as the number of decision variables. However, if the values of $P_{\text{bc},t}$, $P_{\text{bd},t}$, $P_{\text{el},t}$, $P_{\text{fc},t}$, and $P_{\text{sp},i,t}$ are decided, other variables can be derived according to \eqref{f_22} and \eqref{f_23}. Since $P_{\text{bc},t}$ and $P_{\text{bd},t}$ must satisfy \eqref{f_7}, a single variable can be used to represent them simultaneously. Similarly, a single variable is needed to represent $P_{\text{el},t}$ and $P_{\text{fc},t}$ simultaneously. In summary, $2+J$ variables should be decided in each time slot $t$ and $2+J$ agents are considered in this paper. In the following parts, the states, actions, and reward functions related to BESS agent, Thermal-load agents, and HESS agent are designed, respectively.

\subsection{BESS agent}
\subsubsection{Environment State}
According to \eqref{f_22},~\eqref{f_24}-\eqref{f_26}, the calculations of $C_{1,t}$, $C_{2,t}$, and $C_{3,t}$ are related to $v_{t}/\tau_{t}$,~ $P_{\text{pv},t}$,~$P_{\text{load},t}$,~$\mu_{\text{e},t}$,~$B_t$. Since $\tau_{t}$ is often related to $v_{t}$\cite{YuJIOT2020} or is a constant\cite{Ye2020}, $\tau_{t}$ is not adopted as a part of state vector for simplicity. Thus, environment state related to BESS agent can be designed by $s_{\text{b},t}=(v_{t},P_{\text{pv},t},P_{\text{load},t},\mu_{\text{e},t},B_t, t)$.

\subsubsection{Action}
According to \eqref{f_7}, simultaneous BESS charging and discharging are not allowed. Let $a_{b,t}$ be the charging/discharging power of BESS. Then, we have $P_{\text{bc},t}=a_{\text{b},t}$ and $P_{\text{bd},t}=0$ if $a_{\text{b},t} > 0$. Otherwise, $P_{\text{bc},t}=0$ and $P_{\text{bd},t}=a_{\text{b},t}$. Therefore, the constraints \eqref{f_5}-\eqref{f_6} could be guaranteed. To guarantee the feasibility of \eqref{f_4}, we have $0 \le P_{\text{bc},t} \le \min \{P^{\max}_{\text{bc}},~\frac{B^{\max}-B_t}{\eta_{\text{bc}}\Delta t}\} $ if $a_{\text{b},t} >0$. Similarly, $\max \{-P_{\text{bd}}^{\max},\frac{(B^{\min}-B_t)\eta_{\text{bd}}}{\Delta t}\}  \le P_{\text{bd},t} \le 0$ if $a_{\text{b},t} <0$. Based on the above description, $a_{\text{b},t}$ is selected as the action of the BESS agent.

\subsubsection{Reward}
According to \eqref{f_26}, it can be known that the reward of the BESS agent is related to $C_{3,t}$. Moreover, to promote the coordination between BESS agent and HESS agent, the common costs $C_{1,t}$ and $C_{2,t}$ should be considered in the reward design of the BESS agent. For simplicity, the same penalty related to the common costs is imposed on BESS agent and HESS agent. As a result, the reward of BESS agent can be designed as follows, i.e., $r_{\text{b},t}=-(\frac{C_{1,t}+C_{2,t}}{2}+C_{3,t})$.

\subsection{Thermal-load agents}
\subsubsection{Environment State}
According to \eqref{f_19}, the temperature deviation should be penalized on each agent so that comfortable range can be maintained. Moreover, to promote the coordination among all thermal-load agents and HESS agent, $C_{5,t}$ and $C_{6,t}$ should be considered in the reward design of each thermal-load agent and HESS agent. Since the temperature deviation, $C_{5,t}$, and $C_{6,t}$ depend on $\beta_{\text{in},i,t}$,~$\beta_{\text{out},t}$,~$\lambda_{g,t}$, and $Q_{\text{th},t}$, the environment state of $i$-th agent can be designed as follows, i.e., $s_{\text{th},i,t}=(Q_{\text{th},t},~\beta_{\text{in},i,t},~\beta_{\text{out},t},~\lambda_{\text{g},t},~t)$.

\subsubsection{Action}
Since each thermal-load agent needs to make a decision on $P_{\text{sp},i,t}$, the action of $i$-th thermal-load agent can be designed by $a_{\text{th},i,t}=P_{\text{sp},i,t}$. To speed up the learning of agents, the following rules are adopted, i.e.,
\begin{equation}\label{f_31}
{a_{th,i,t}} = \left\{ \begin{array}{l}
0,~~~\text{if}~{\beta _{{\rm{in}},i,t}} \le \beta _i^{\min }~\text{or}~ {\beta _{{\rm{out}},t}} \le \beta _i^{\max }\\
{P_{{\rm{sp}},i,t}},~~~\text{otherwise}.
\end{array} \right.
\end{equation}

\subsubsection{Reward}
As mentioned in the descriptions related to state design, the reward of $i$-th agent consists of three parts, i.e., the penalties imposed on temperature deviation, CWT depreciation cost, and gas purchasing cost. Therefore, the reward of each agent $i$ can be designed as follows, i.e., $r_{\text{th},t}=-(\frac{C_{5,t}+C_{6,t}}{J+1}+\varpi_{i,t})$, where $\varpi_{i,t}=\pi_{\text{th}}({\left[ {{\beta_{\text{in},t+1}} - {\beta^{\max }}} \right]^ + } + {\left[ {{\beta^{\min}} - {\beta_{\text{in},t+1}}} \right]^+}$, and $\pi_{\text{th}}$ denotes a positive penalty coefficient.

\subsection{HESS agent}
\subsubsection{Environment State}
According to \eqref{f_27}, the calculation of $C_{4,t}$ is related to $I_{\text{el},t-1}^{\text{on}}$ and $I_{\text{fc},t-1}^{\text{on}}$. In addition, to promote the coordination among HESS agent, BESS agent, and thermal-load agents, the rewards of BESS agent and thermal-load agents should be used for the reward shaping of HESS agent. Therefore, the state of HESS agent can be described by $s_{\text{h},t}=(I_{\text{el},t-1}^{\text{on}},I_{\text{fc},t-1}^{\text{on}},v_{t},B_t,H_t,P_{\text{pv},t},P_{\text{load},t},\mu_{\text{e},t},Q_{\text{th},t},\beta_{\text{out},t},\lambda_{\text{g},t},\\ \beta_{\text{in},i,t}|_{1\leq i\leq J},t)$.

\subsubsection{Action}
According to \eqref{f_15}, the operations of electrolyzer and fuel cell can not happen simultaneously. For simplicity, $a_{\text{h},t}$ is adopted to represent its operation. According to \eqref{f_16}-\eqref{f_17}, we have $P_{\text{fc},t}$=0, $P_{\text{el},t} \in (0,~P_{\text{el}}^{\max}]$ if $a_{\text{h},t}>0$. Otherwise, we have $P_{\text{el},t}$=0, $P_{\text{fc},t} \in [-P_{\text{fc}}^{\max},~0]$. To guarantee the feasibility of \eqref{f_14}, we have $0 \le P_{\text{el},t} \le \min \{P^{\max}_{\text{el}},~\frac{H^{\max}-H_t}{\omega_{\text{el}}\Delta t}\} $ if $a_{\text{h},t} >0$. Similarly, $\max \{-P_{\text{fc}}^{\max},\frac{(H^{\min}-H_t)\omega_{\text{fc}}}{\Delta t}\}  \le P_{\text{fc},t} \le 0$ if $a_{\text{h},t} <0$.

\subsubsection{Reward}
According to the descriptions in state design, the reward of the HESS agent consists of several components related to $C_{1,t}$, $C_{2,t}$, $C_{5,t}$, and $C_{6,t}$. In addition, to reduce the waste of thermal energy generated by fuel cell, a penalty is imposed on the HESS agent, i.e., $\xi_t=\pi_{\text{fc}}(Q_{\text{fc},t}\eta_{\text{h2c}}-(P_{\text{tc},t}+P_{\text{td},t}+\sum_{i=1}^{N_b}P_{\text{sp},i,t}-P_{\text{gb},t}\eta_{\text{h2c}})\Delta t)$, where $\pi_{\text{fc}}$ is a positive penalty parameter. In summary, the reward of the HESS agent is design by $r_{\text{h},t}=-(\frac{C_{1,t}+C_{2,t}}{2}+C_{4,t}+\frac{C_{5,t}+C_{6,t}}{J+1}+\xi_t)$.

\textbf{Remark~1:} After the actions of thermal-load agents and HESS agent are taken, the actions of gas boiler and CWT can be decided accordingly. To be specific, when $Q_{\text{fc},t}\eta_{\text{h2c}}>\sum_{i=1}^JP_{\text{sp},i,t}\Delta t$, CWT will operate in charging mode (i.e., $P_{\text{td},t}=0$) and the thermal power input $P_{\text{tc},t}$ is $\min(\frac{Q_{\text{fc},t}\eta_{\text{h2c}}}{\Delta t}-\sum_{i=1}^JP_{\text{sp},i,t},P_{\text{tc}}^{\max},\frac{Q_{\text{th}}^{\max}-Q_{\text{th},t}}{\eta_{\text{tc}}\Delta t})$. Under this situation, $P_{\text{gb},t}=0$. When $Q_{\text{fc},t}\eta_{\text{h2c}}\leq \sum_{i=1}^J P_{\text{sp},t}\Delta t$, CWT will operate in discharging mode (i.e., $P_{\text{tc},t}=0$) and the thermal output is $\min(\sum_{i=1}^JP_{\text{sp},i,t}\Delta t-Q_{\text{fc},t}\eta_{\text{h2c}},P_{\text{td}}^{\max}\Delta t,Q_{\text{th},t}\eta_{\text{td}})$. Under this situation, $P_{\text{gb},t}=\min(\sum_{i=1}^J P_{\text{sp},i,t}-\frac{Q_{\text{fc},t}\eta_{\text{h2c}}}{\Delta t}-P_{\text{td},t},P^{\max}_{\text{gb}})$. Consequently, the total thermal supply power to buildings is $P_{\text{thermal},t}=P_{\text{gb},t}+P_{\text{td},t}+\frac{Q_{\text{fc},t}}{\eta_{\text{h2c}}\Delta t}$. When $\sum_{i=1}^JP_{\text{sp},i,t}>P_{\text{thermal},t}$, the actual thermal input of building $i$ is decided by $P_{\text{thermal},t}\frac{P_{\text{sp},i,t}}{\sum_{i=1}^JP_{\text{sp},i,t}}$.

\textbf{Remark~2:} Since continuous action spaces have infinite number of solutions, we divide action spaces of BESS, HESS, and thermal load agents into multiple discrete values for the purpose of simplifying the solving of Markov game. Taking BESS for example, the action of BESS agent has $N_{\text{bess}}$ choices, i.e., $a_{\text{b},t}\in \{ \mathcal{A}_{\text{bess},1},\cdots,\mathcal{A}_{\text{bess},N_{\text{bess}}}\}$, where $\mathcal{A}_{\text{bess},1}=-P^{\max}_{\text{bd}}$, $\mathcal{A}_{\text{bess},N_{\text{bess}}}=P^{\max}_{\text{bc}}$, $\mathcal{A}_{\text{bess},n_b}-\mathcal{A}_{\text{bess},n_b-1}=(P^{\max}_{\text{bd}}+P^{\max}_{\text{bc}})/(N_{\text{bess}}-1)$, $2\leq n_b \leq N_{\text{bess}}$. Similarly, we have $a_{\text{h},t}\in \{ \mathcal{A}_{\text{hess},1},\cdots,\mathcal{A}_{bess,N_{\text{hess}}}\}$, where $\mathcal{A}_{\text{hess},1}=-P^{\max}_{\text{fc}}$, $\mathcal{A}_{\text{hess},N_{\text{hess}}}=P^{\max}_{\text{el}}$, $\mathcal{A}_{\text{hess},n_h}-\mathcal{A}_{\text{hess},n_h-1}=(P^{\max}_{\text{fc}}+P^{\max}_{\text{el}})/(N_{\text{hess}}-1)$, $2\leq n_h \leq N_{\text{hess}}$. $a_{th,i,t}\in
\{0, \frac{P_{\text{sp},i}^{\max}}{N_{\text{thermal}}-1},\cdots,\frac{P_{\text{sp},i}^{\max}(N_{\text{thermal}}-2)}{N_{\text{thermal}}-1},P_{\text{sp},i}^{\max}\}$. Note that when some actions are not feasible, they should be adjusted as introduced in next section. At this time, actions may be continuous.

\section{The Proposed Energy Management Algorithm}\label{s4}

In this section, we propose an energy management algorithm to solve Markov game in Section~\ref{s3} based on MADACR. To be specific, some reasonable rules are used to adjust actions generated by actor networks so that the selected actions are feasible and the exploration space could be reduced for better convergence performance. Moreover, multi-agent discrete actor-critic with the framework of centralized training and decentralized execution is used in algorithm design. In the following parts, we first introduce the key idea of the proposed algorithm. Then, we describe the algorithm details.

\begin{figure}[!htb]
\centering
\includegraphics[scale=1.15]{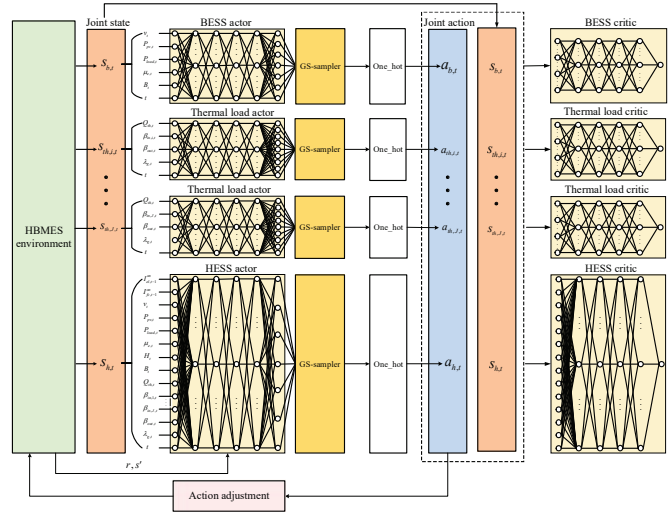}
\caption{The overall framework of the proposed algorithm.}\label{fig_2}
\end{figure}

\subsection{The key idea of the proposed algorithm}\label{s41}
The overall framework of the proposed algorithm is illustrated by Fig.~\ref{fig_2}, where $J+2$ agents can be identified, i.e., a BESS agent, an HESS agent, and $J$ thermal load agents. Note that each agent consists of one actor and one critic, where the actor is used by each agent to generate action $a_i$ under a given state $s_i$. Then, the joint action of all agents $a=(a_{\text{b}},a_{\text{th},1},\cdots,a_{\text{th},J},a_{\text{h}})=(a_1,\cdots,a_i,\cdots,a_{J+2})$ is adjusted according to some rules introduced in the next subsection. Consequently, that joint action is executed and the environment returns a new sate $s'=(s_{\text{b}},s_{\text{th},1},\cdots,s_{\text{th},J},s_{\text{h}})=(s_1,\cdots,s_i,\cdots,s_{J+2})$ and a reward vector $r$. Then, experience transition tuple ($s,a,s',r$) is stored in an experience memory for training. The function of a critic is to evaluate the action-value function $Q_i(s_i,a_i)$ under the given state $s_i$ and action $a_i$. When computing the $Q_i(s_i,a_i)$, the states and actions of other agents are also incorporated to ensure that the training environment is stationary\cite{NguyenMAS2019}. In Fig.~\ref{fig_2}, it can be observed that all actors and critics are represented by deep neural networks, which are composed of one input layer, multiple hidden layers and one output layer. To support discrete actions, Gumbel-Softmax (GS) sampler is adopted to approximate categorical sampler. Above all, GS sampler can support back-propagation when computing policy gradient with respect to actor parameters. In contrast, that gradient under categorical sampler is non-differentiable and can not be back-propagated \cite{Jang2017}.

When training multiple agents, some techniques (e.g., target actors/critics and experience replay) are adopted to stabilize the training process. To be specific, the network parameters of each critic in agent $i$ are updated according to the loss function related to temporal-difference error, i.e.,
\begin{equation}\label{f_32}
L_{i,c} = \mathbb{E}_{s,a,s',r}[y_i-Q_i(s,a)]^2,
\end{equation}
where the target value $y_i$ is calculated by
\begin{equation}\label{f_33}
y_i = r_i+ \gamma \bar{Q}_i(s',a'_1,\cdots,a'_{j},\cdots,a'_{J+2}),
\end{equation}
where $a'_{j}$=$\text{one\_hot}(\arg\max\limits_{j'}(\bar{\mu}_j(a_{j'}|s'_{j'})))$. Here, $\text{one\_hot}(j)$ is a vector with only one non-zero element. Moreover, its index is $j$ and value is 1. $\bar{Q}_i$ and $\bar{\mu}_i$ denotes the target critic and target actor related to agent $i$, respectively.

Once the parameter adjustment of critics is finished, the actor parameters in each agent $i$ will be updated by maximizing the policy gradient as follows,
\begin{equation}\label{f_34}
\nabla_{\vartheta_i} \Upsilon=\mathbb{E}_{s\sim S^{\mu}}[\nabla_{\vartheta_i} Q_i(s,a^*)|_{a_i^*=\Psi_i, a_{j\neq i}^*=\Psi_j}],
\end{equation}
where the actions $\Psi_i=\text{one\_hot}(\arg\max\limits_{i}(GS(\mu_{\vartheta_i}(a_i|s_i))))$ and
$\Psi_j=\text{one\_hot}(\arg\max\limits_{j}(\mu_{\vartheta_j}(a_j|s_j)))$. Here, $GS(\cdot)$ denotes the output of GS sampler. Note that $\text{one\_hot}(\cdot)$ is used to ensure that only one discrete action can be selected so that the action is meaningful in practice, while $GS(\cdot)$ is adopted to approximate the discrete action and ensure that the gradient with respect to $\vartheta_i$ used in back-propagation is differentiable.

\subsection{Algorithmic Details}\label{s42}
The proposed algorithm consists of training algorithm (i.e., Algorithm~\ref{alg_1}) and execution algorithm (i.e., Algorithm~\ref{alg_2}). Next, we will describe the details of two algorithms.

In Algorithm~\ref{alg_1}, a replay memory $\mathcal{D}$ is initialized in line 1. Then, the preprocessing function $\phi(s)$ is introduced to normalize the environment state $s_{t}$ as in \cite{YuJIOT2020}, which can facilitate the learning process of the propose algorithm. In lines 3-4, we initialize the weight parameters of actors/critics and target actors/critics, which have the same network architecture. In each time slot $t$ of an episode, each agent $i$ takes an action $a_i$ in parallel based on its local state $s_i$, i.e.,
\begin{equation}\label{f_35}
a_i=\text{one\_hot}(\arg\max\limits_{i}(GS(\mu_i(a_i|\phi(s_i))))).
\end{equation}

To ensure that the selected discrete actions are feasible and reduce exploration spaces for better convergence performances, some reasonable rules are adopted as follows: (1) Discrete actions of HESS and BESS agents should be adjusted to meet their respective physical constraints defined in Section~\ref{s3}-A and Section~\ref{s3}-C; (2) Discrete actions of thermal load agents should be adjusted according to \eqref{f_31}; (3) If the event $P_{\text{pv},t}>P_{\text{load},t}$ happens frequently during the periods with high prices and the selling price is lower than any buying price, an reasonable way is to store rather than sell the excess energy. The reason is that the former can result in a reduced quantity of buying electricity when $P_{\text{pv},t}<P_{\text{load},t}$ happens in future. Since any buying price is higher than selling price, the benefit of the former is higher than that of the latter. Due to high startup cost of HESS, BESS has higher priority of charging electricity than HESS. In other words, $0 \le P_{\text{bc},t} \le \min \{P_{\text{pv},t}-P_{\text{load},t},P^{\max}_{\text{bc}},~\frac{B^{\max}-B_t}{\eta_{\text{bc}}\Delta t}\}$ when $P_{\text{pv},t}>P_{\text{load},t}$. Let $\zeta_t=\max(0,P_{\text{pv},t}-P_{\text{load},t}-P_{\text{bc},t})$. Then, we have $0 \le P_{\text{el},t} \le \min \{\zeta_t,P^{\max}_{\text{el}},~\frac{H^{\max}-H_t}{\omega_{\text{el}}\Delta t}\}$. Similarly, when the stored energy in BESS and HESS are discharged in future when $P_{\text{pv},t}<P_{\text{load},t}$, BESS has higher priority of discharging electricity than HESS.

After receiving the joint action of all agents, the environment returns a new state $a'$ and a reward $r$. Next, the experience transition tuple $(\phi(s),a,\phi(s'),r)$ is stored in the memory $\mathcal{D}$. When the number of transition tuples $M_{\text{size}}$ exceeds $N_m$, the multi-agent training process would be triggered. However, for the purpose of stabilizing learning process, the training frequency is decreased by adopting another condition, i.e., mod($ep$,$T_{\text{fre}}$)=0, where $ep$ denotes the episode index and $T_{\text{fre}}$ means that training is conducted every $T_{\text{fre}}$ episodes. In lines 14-17, each agent update its actor and critic parameters based on the sampled mini-batch data with $K$ transition tuples. Then, the loss function used for updating critic parameters $\theta_i$ can be calculated by
\begin{equation}\label{f_36}
L_{\theta_i,c} = {1\over K}\sum\nolimits_{k=1}^{K}[y_i^{k}-Q_{\theta_i}(\phi(s^{k}),a^{k})]^2,
\end{equation}
where $y_i^{k} = r_i^{k}+ \gamma \bar{Q}_{\theta_i}(\phi(s'^{k}),a'_1,\cdots,a'_{j},\cdots,a'_{J+2})$, $a'_{j}=\text{one\_hot}(\arg\max\limits_{j'}(\bar{\mu}_j(a_{j'}|\phi(s_{j'}^k))))$.

Similarly, the policy gradient used for updating actor parameters can be calculated by
\begin{equation}\label{f_37}
\nabla_{\vartheta_i} \Upsilon={1\over K}\sum\limits_{k=1}^{K}\Big[\nabla_{\vartheta_i} Q_i(\phi(s^k),a^*)|_{a_i^*=\Psi_i^k, a_{j\neq i}^*=\Psi_j^k}\Big],
\end{equation}
where $\Psi_i^k=\text{one\_hot}(\arg\max\limits_{i}(GS(\mu_{\vartheta_i}(a_i|\phi(s_i^k)))))$,
$\Psi_j^k=\text{one\_hot}(\arg\max\limits_{j}(\mu_{\vartheta_j}(a_j|\phi(s_j^k))))$.

After updating parameters of actors and critics, target actors and critics will be adjusted as shown in line 19. Note that once the above-mentioned training process is finished, the obtained actor networks could be used for real-time decisions as shown in Algorithm 2. In each time slot $t$, each agent $i$ takes an action $a_{i,t}$ based on the local observation $s_{i,t}$ in parallel, which is shown in line 3. Then, the joint action of all agents will be adjusted and executed. As a result, a new state is returned. The above process repeats until the end of testing period. Since just current observation and forward propagation of deep neural networks are involved in execution algorithm, the proposed energy management algorithm has low computational complexity and does not require any prior knowledge of uncertain parameters and explicit building thermal dynamics models.

\begin{algorithm}[h]
\caption{Training Algorithm for HBMES Energy Management}
\label{alg_1}
\LinesNumbered
Initialize replay memory $\mathcal{D}$ with size $N_m$\;

Initialize preprocess function $\phi(s)$\;

Randomly initialize critic networks $Q_i(\phi(s),a)$ and actor network $\mu_i(a_i|\phi(s))$ with weights $\theta_i$, $\vartheta_i$, respectively.

Initialize target networks $\bar{Q}_i$ and $\bar{\mu}_i$ by copying: $\bar{\theta}_i\Leftarrow\theta_i$, $\bar{\vartheta}_i\Leftarrow\vartheta_i$

\For{$ep$=1,~2,~$\cdots$,~$M$}
{
    Receive the initial environment state $s$

    \For{$t$=0,~1,~$\cdots$,~$T$-1}
    {

        Each agent $i$ selects an action according to \eqref{f_35} in parallel\;

        Rule-based action adjustment\;

        Execute action $a$ and obtain next state $s'$ and reward $r$ from the environment\;

        Store $(\phi(s),a, r,\phi(s'))$ in $\mathcal{D}$\;

        $s\leftarrow s'$\;

        \If{$M_{\text{size}}\geq N_m$ and mod($\text{ep}$,$T_{\text{fre}}$)=0}
        {

              \For{agent~$i$=1,~$\cdots$,~$J+2$}
              {
                 Sample a mini-batch of $K$ transitions $(\phi(s^{k}),a^{k},r^{k},\phi(s'^{k}))$ from $\mathcal{D}$\;

                 Update critic network by minimizing the loss function in \eqref{f_36}\;

                 Update actor network by maximizing the policy gradient in \eqref{f_37}\;
              }

          Update target network parameters for each agent $i$: $\bar{\theta}_i \leftarrow \rho \theta_i+(1-\rho)\bar{\theta}_i$, $\bar{\vartheta}_i \leftarrow \rho \vartheta_i+(1-\rho)\bar{\vartheta}_i$\;

        }

    }
}
\end{algorithm}

\begin{algorithm}[h]
\caption{Execution Algorithm for HBMES Energy Management}
\label{alg_2}
\setcounter{AlgoLine}{0}
\LinesNumbered
\KwIn{Actor networks $\mu$ with weights $\vartheta$}
\KwOut{Actions: $a_{i,t}$}

Receive initial environment state $s_{0}=(s_{1,0},\cdots,s_{J+2,0})$

\For{$t$=0,~1,~$\cdots$,~$T_{\text{test}}$-1}
{
     Each agent selects an action $a_{i,t} = \text{one\_hot}(\arg\max\limits_{i}(\mu_i(a_i|\phi(s_{i,t}))))$ in parallel\;

     Rule-based action adjustment\;

     Execute the action $a_{t}=(a_{1,t},\cdots,a_{i,t},\cdots,a_{J+2,t})$\;

     Obtain next state $s_{t+1}$ from the environment\;

}
\end{algorithm}

%
%

\section{Performance Evaluation}\label{s5}
In this section, we evaluate the performance of the proposed algorithm. To be specific, we first describe the simulation setup. Then, four benchmarks are adopted for performance comparisons. Finally, we provide some simulation results about algorithmic convergence, algorithmic effectiveness, and algorithmic robustness under varying thermal disturbances.

\subsection{Simulation setup}\label{s51}
Real-world traces related to electricity price, power load, PV generation, and outdoor temperature are adopted in simulations. To be specific, retail commercial price during June 1 and Sept. 30 of 2019 in Beijing is used\footnote{http://fgw.beijing.gov.cn/}. Moreover, power demand and outdoor temperature data from Pecan Street database\footnote{https://www.pecanstreet.org/} are used. Furthermore, we use solar irradiance data during June 1 and Sept. 30 of 2019 from NREL Solar Radiation Research Laboratory\footnote{https://midcdmz.nrel.gov/}. In these traces, the data within 90 days and 30 days are used for training and testing, respectively. Note that main simulation parameters are summarized in Table~\ref{table_1} and all simulations are conducted on a desktop computer with Intel Core(TM) i9-9900 CPU and 64GB RAM. To simulate the thermal load, cooling mode is considered and the following building thermal dynamics model $\mathcal{F}$ is adopted for simplicity, i.e., $\beta_{\text{in},i,t+1}=\varepsilon_{\text{hvac}}\beta_{\text{in},i,t}+(1-\varepsilon_{\text{hvac}})(\beta_{\text{out},t}-P_{\text{sp},i,t}\eta_{\text{hvac}}/A_i)+\varrho_{i,t}$\cite{YuJIOT2020}, and the performance of the proposed algorithm under different $\varrho_{i,t}$ will be discussed later. Similar to \cite{Ye2020}, selling price is fixed to be 0.1 RMB/kWh and smaller than any buying price to avoid any arbitrage behavior, i.e., buying electricity at low prices and selling electricity at high prices. To evaluate the performance of the proposed algorithm under different PV generation modes, two cases are considered, i.e., the total radiation area of solar panels is $h_{\text{pv}}$=$100\text{m}^2$ under case-1 and $h_{\text{pv}}$=$250\text{m}^2$ under case-2.

\begin{table}[!htbp]
\center
\caption{Main Parameter Settings}\centering\label{table_1}
\renewcommand{\baselinestretch}{2.0}
\begin{tabular}{|p{0.95\columnwidth}|}
\hline
\makecell[c]{\textbf{PV generation, gas boiler, and carbon emission}}\\
\hline
$\eta_{\text{pv}}$=$0.2$\cite{LiangTSG2019}, $\eta_{\text{gb}}$=0.95\cite{Ye2020}, $\lambda_{\text{gb}}$=0.287\text{RMB/kWh}\cite{Ye2020},
$\mu_{\text{e},t}$=0.968\text{kg/kWh}, $\mu_\text{c}$=0.06\text{RMB/kg}\cite{Dong2020}, $P^{\max}_{\text{gb}}$=20kW \\
\hline
\makecell[c]{\textbf{BESS}}\\
\hline
$B^{\text{min}}$=0\text{kWh}, $B^{\text{init}}$=0\text{kWh}, $B^{\text{max}}$=40\text{kWh}, $P_{\text{bc}}^{\max}$=20kW,
$P_{\text{bd}}^{\max}$=30\text{kW}, $\eta_{\text{bc}}$=$\eta_{\text{bd}}$=0.95\cite{YuJIOT2020}, $\psi_{\text{BESS}}$=0.001RMB/kW\cite{YuJIOT2020}\\
\hline
\makecell[c]{\textbf{CWT}}\\
\hline
$\eta_{\text{tc}}$=0.9, $\eta_{\text{td}}=0.9$, $Q_{\text{th}}^{\text{max}}$=50\text{kWh}, $Q_{\text{th}}^{\text{init}}$=0\text{kWh},
$P_{\text{tc}}^{\max}$=10\text{kWh},
$P_{\text{td}}^{\max}$=10\text{kWh}, $\psi_{\text{CWT}}$=0.005RMB/kW\\
\hline
\makecell[c]{\textbf{HESS}}\\
\hline
$\omega_{\text{fc}}$=0.2397Nm$^3$/kWh, $\omega_{\text{el}}$=1.4985kWh/Nm$^3$, $\eta_{\text{hr}}$=0.7\cite{Liu2021},
$\eta_{\text{h2e}}$=1.4\cite{Liu2021}, $\eta_{\text{h2c}}$=0.7\cite{Liu2021}, $P_{\text{el}}^{\max}$=20kW, $P_{\text{fc}}^{\max}$=20kW, $\delta_{\text{el}}^{\text{on}}$=0.158\text{RMB}\cite{Garcia2015},
$\delta_{\text{el}}^{\text{su}}$=0.97\text{RMB}\cite{Garcia2015}, $\delta_{\text{el}}^{\text{sd}}$=0.049\text{RMB}\cite{Garcia2015}, $\delta_{\text{fc}}^{\text{on}}$=0.079\text{RMB}\cite{Garcia2015},
$\delta_{\text{fc}}^{\text{su}}$=$\delta_{\text{fc}}^{\text{sd}}$=0.0004\text{RMB}~\cite{Garcia2015}, $H^{\max}$=30\text{Nm$^3$}, $\pi_{\text{fc}}=1$\\
\hline
\makecell[c]{\textbf{Thermal load}}\\
\hline
J=4, $\beta^{\text{init}}$=[21, 20, 22, 21.5]$^{\circ}$\text{C}, $\beta_i^{\text{min}}$=20$^{\circ}$\text{C}, $\beta_i^{\text{max}}$=25$^{\circ}$\text{C}, $\eta_{\text{hvac}}$=2.5, $A=0.5\text{kW}/^\circ$F, $\varepsilon_{\text{hvac}}=0.8$, $P_{\text{sp}}^{\max}$=20kW\\
\hline
\makecell[c]{\textbf{Training algorithm related}}\\
\hline
$\gamma$=0.95,$N_{a}^{h}$=$N_{c}^{h}$=\{128,128,128\},$\mathcal{D}$=120000,$M$=30000,~$\pi_{\text{th}}$=0.35RMB/$^o$F,
$\upsilon_{\text{a}}$=$\upsilon_{\text{c}}$=0.00008, $T$=24, $\Delta t$=1h, $\varrho_{t}$=0, $K$=256, $\rho$=0.001, $T_{\text{test}}$=720, $T_{\text{fre}}$=5, $N_{\text{hess}}=N_{\text{bess}}=7$ (case-1), $N_{\text{hess}}=N_{\text{bess}}=21$ (case-2), $N_{\text{thermal}}=9$ \\
\hline
\end{tabular}
\end{table}

\subsection{Benchmarks}\label{s52}
\begin{itemize}
  \item \textbf{Baseline 1 (B1)}: This scheme controls BESS and HESS using an algorithm similar to \cite{Barbero2020}, i.e., to charge BESS and HESS greedily when there is a surplus of renewable energy and discharge them otherwise. Moreover, this scheme adopts ON/OFF strategy\cite{Wei2017} for building cooling, i.e., $P_{\text{sp},i,t}$=0 if $\beta_{\text{in},i,t}\leq \beta_i^{\min}$ and $P_{\text{sp},i,t}$=$P_{\text{sp},i}^{\max}$ if $\beta_{\text{in},i,t}\geq \beta_i^{\max}$.
  \item \textbf{Baseline 2 (B2)}: This scheme considers the use of BESS by greedily charging electricity at the minimum price and discharging electricity to minimize the amount of buying electricity at the maximum price. Moreover, it adopts ON/OFF strategy for building cooling.
  \item \textbf{Baseline 3 (B3)}: This scheme intends to schedule BESS and HESS jointly based on DDQN algorithm\cite{Hasselt2016}\footnote{When scheduling BESS, HESS, and thermal loads jointly, \textbf{B3} is inefficient since the action space increases exponentially with the increase of $J$.}, which is an improved version of DQN\cite{Barbero2020}. Moreover, ON/OFF strategy is adopted for building cooling.
  \item \textbf{Baseline 4 (B4)}: This scheme intends to solve a linear transformation of \textbf{P1} using CPLEX solver in GAMS with the assumption that perfect information about uncertain parameters and building thermal dynamics model can be known, which can provide a performance upper limit for the proposed algorithm.
\end{itemize}


\subsection{Algorithmic convergence process}\label{s53}

\begin{figure}[!htb]
\centering
\includegraphics[scale=0.41]{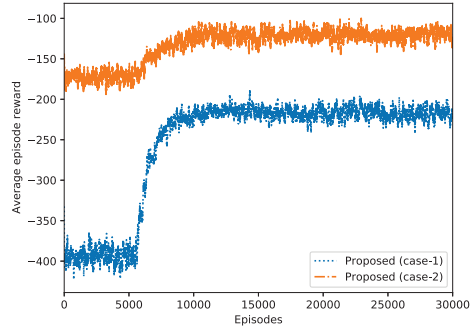}
\caption{Convergence process of the proposed algorithm.}\label{fig_3}
\end{figure}

The convergence process of the proposed algorithm is shown in Fig.~\ref{fig_3}. It can be observed that the average total episode reward of all agents received over the past 50 episodes gradually increases and becomes more and more stable under case-1 and case-2. Due to the existence of exploration incurred by GS-sampler and random parameters, that rewards under two cases still fluctuate within a small range at the end of training period.

\subsection{Algorithmic effectiveness}\label{s54}

\begin{figure*}
\centering
\subfigure[Operational cost]{
\begin{minipage}[b]{0.23\textwidth}
\includegraphics[width=1\textwidth]{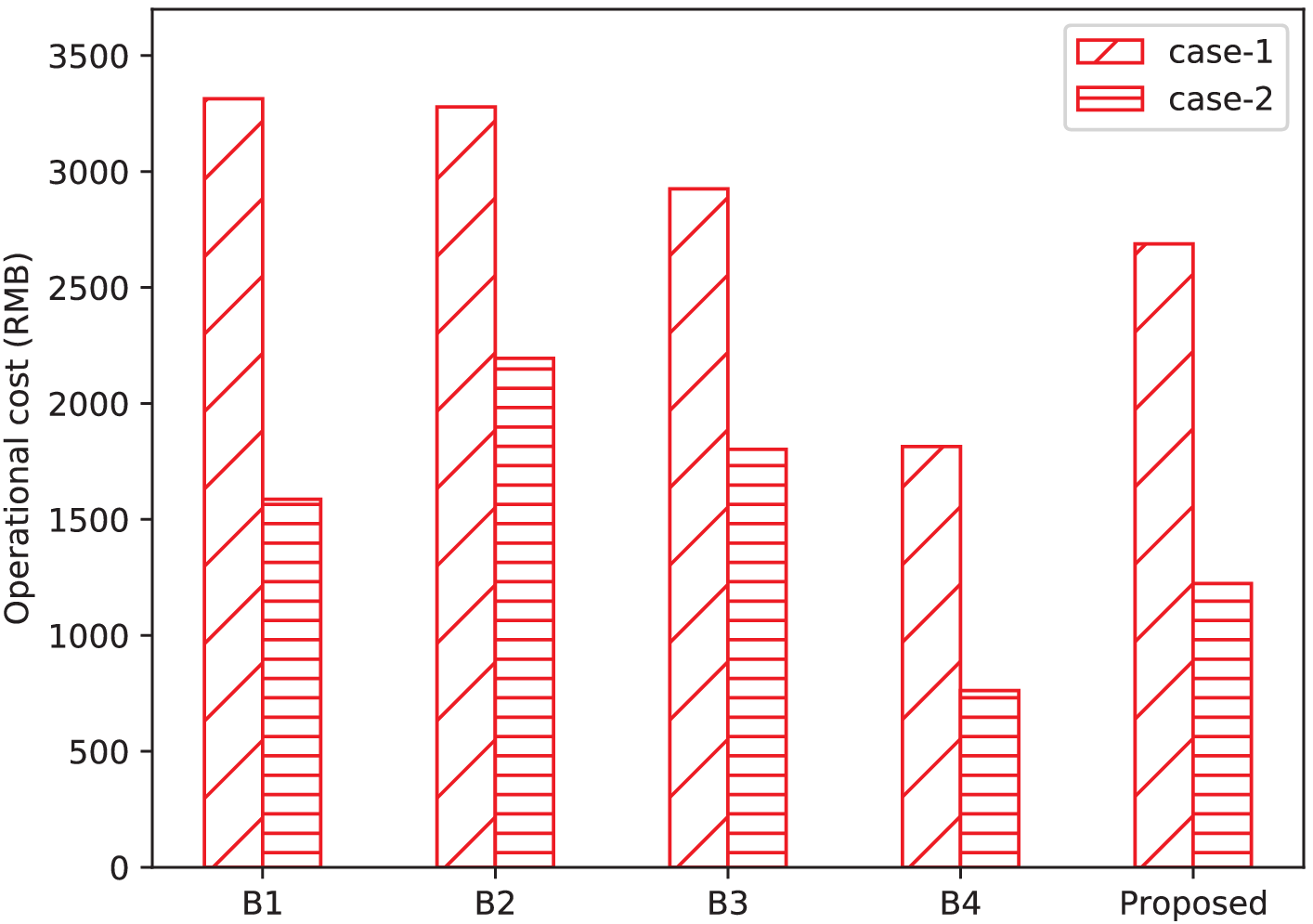}
\end{minipage}}
\subfigure[ATD]{
\begin{minipage}[b]{0.23\textwidth}
\includegraphics[width=1\textwidth]{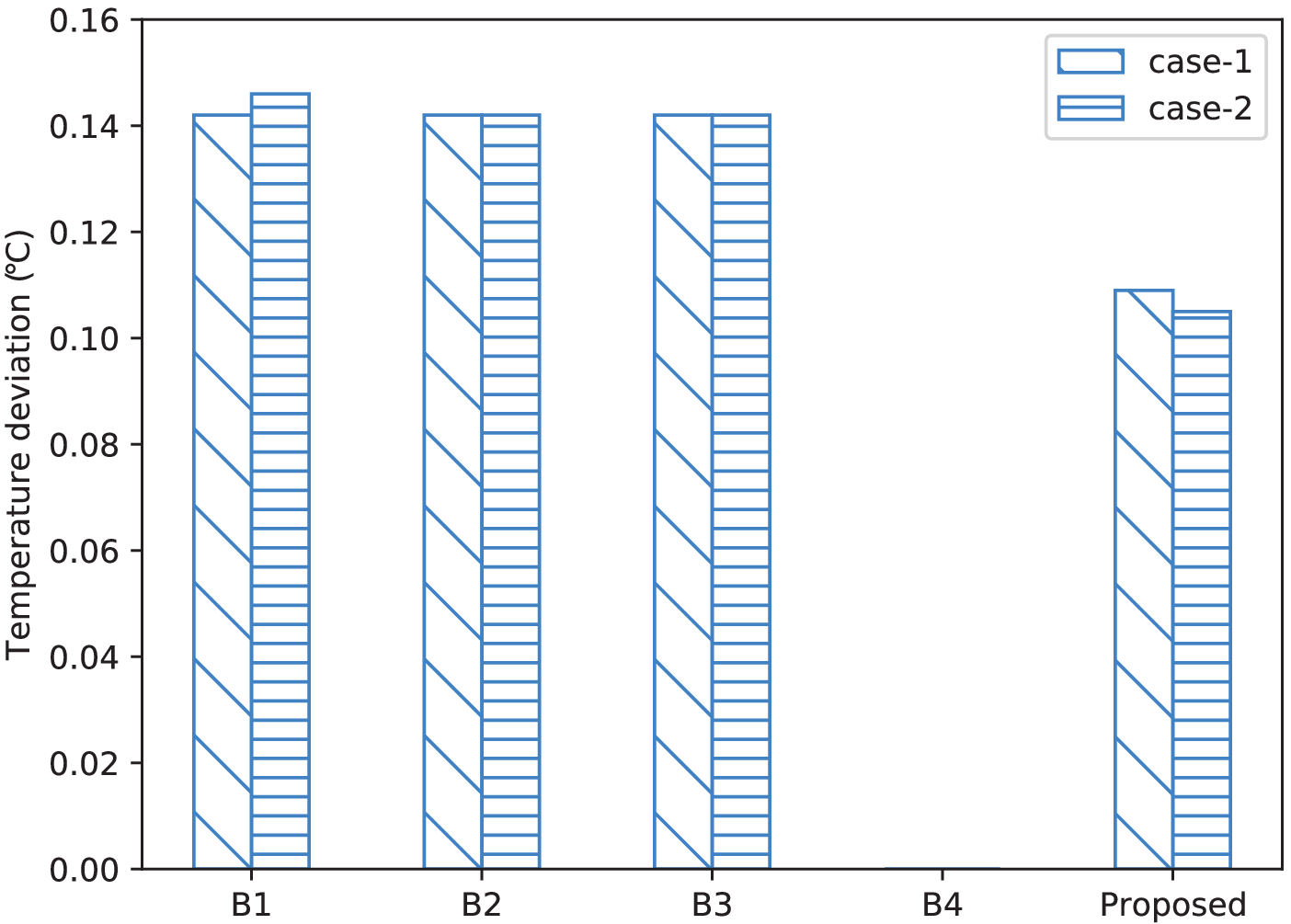}
\end{minipage}}
\subfigure[Cost decomposition (case-1)]{
\begin{minipage}[b]{0.21\textwidth}
\includegraphics[width=1\textwidth]{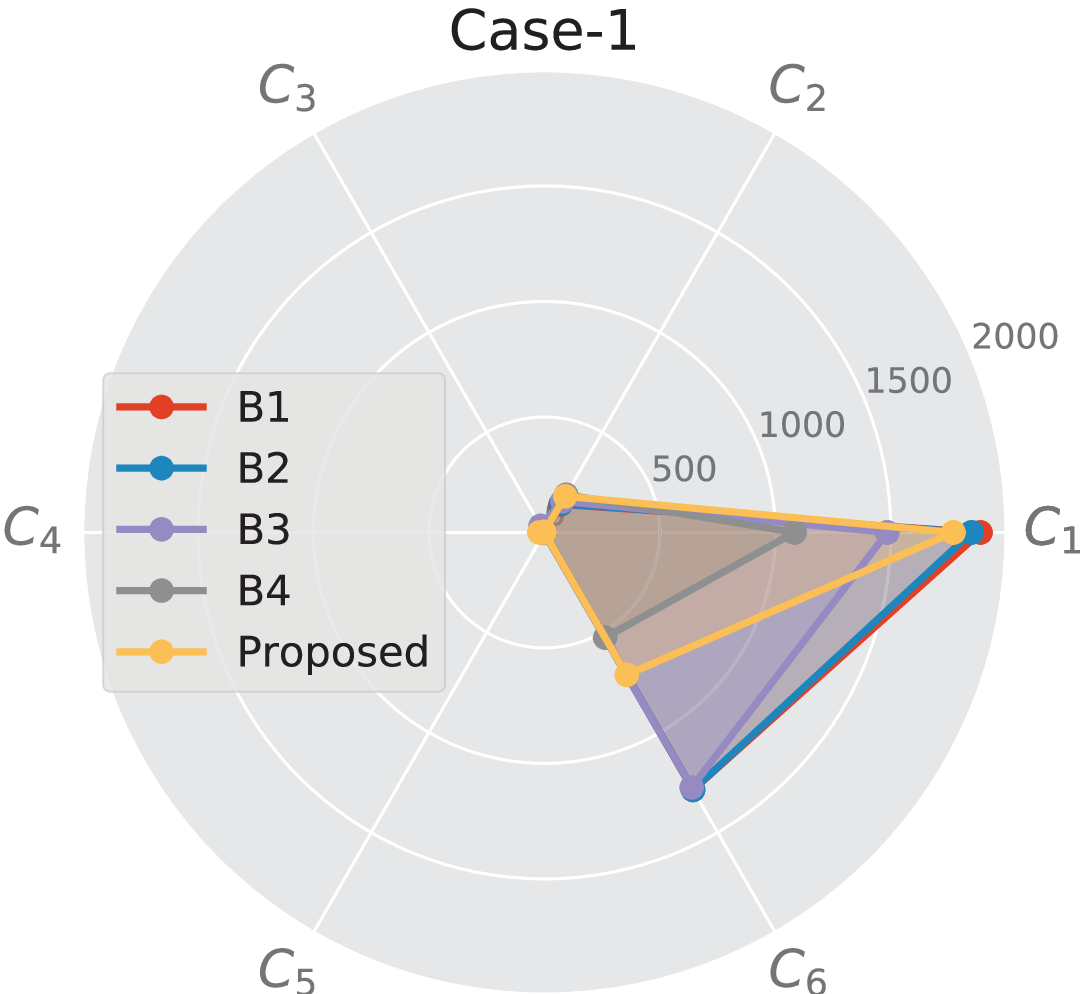}
\end{minipage}}
\subfigure[Cost decomposition (case-2)]{
\begin{minipage}[b]{0.21\textwidth}
\includegraphics[width=1\textwidth]{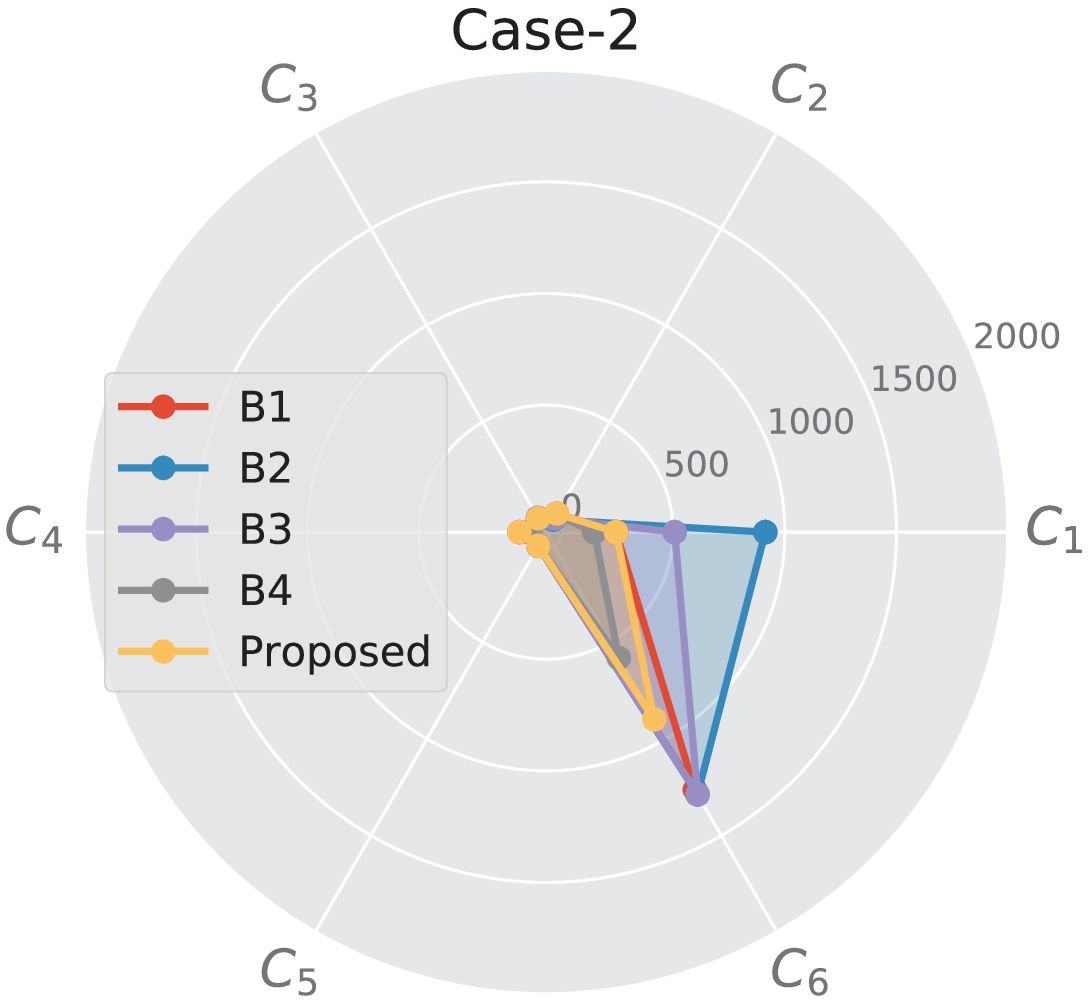}
\end{minipage}}
\caption{Effectiveness of the proposed algorithm.} \label{fig_5}
\end{figure*}

Performance comparisons among all schemes are shown in Fig.~\ref{fig_5}, where operational cost, average temperature deviation (ATD), and operational cost decompositions under case-1 and case-2 can be identified. Compared with \textbf{B1}-\textbf{B3}, the proposed algorithm can reduce operational cost without sacrificing ATD by 18.89\%(22.94\%), 18.02\%(44.26\%), 8.10\%(32.13\%) under case-1 (case-2), respectively. Although \textbf{B4} achieves the best performance, it requires perfect information about random parameters and explicit building thermal dynamics models. Thus, the proposed algorithm is more practical than \textbf{B4}. Although \textbf{B1}, \textbf{B2} and \textbf{B3} use the same strategy for building temperature control, they have slightly different ATDs due to the input power adjustment introduced in \textbf{Remark 1}.

\begin{figure*}
\centering
\subfigure[Price and net load]{
\begin{minipage}[b]{0.23\textwidth}
\includegraphics[width=1\textwidth]{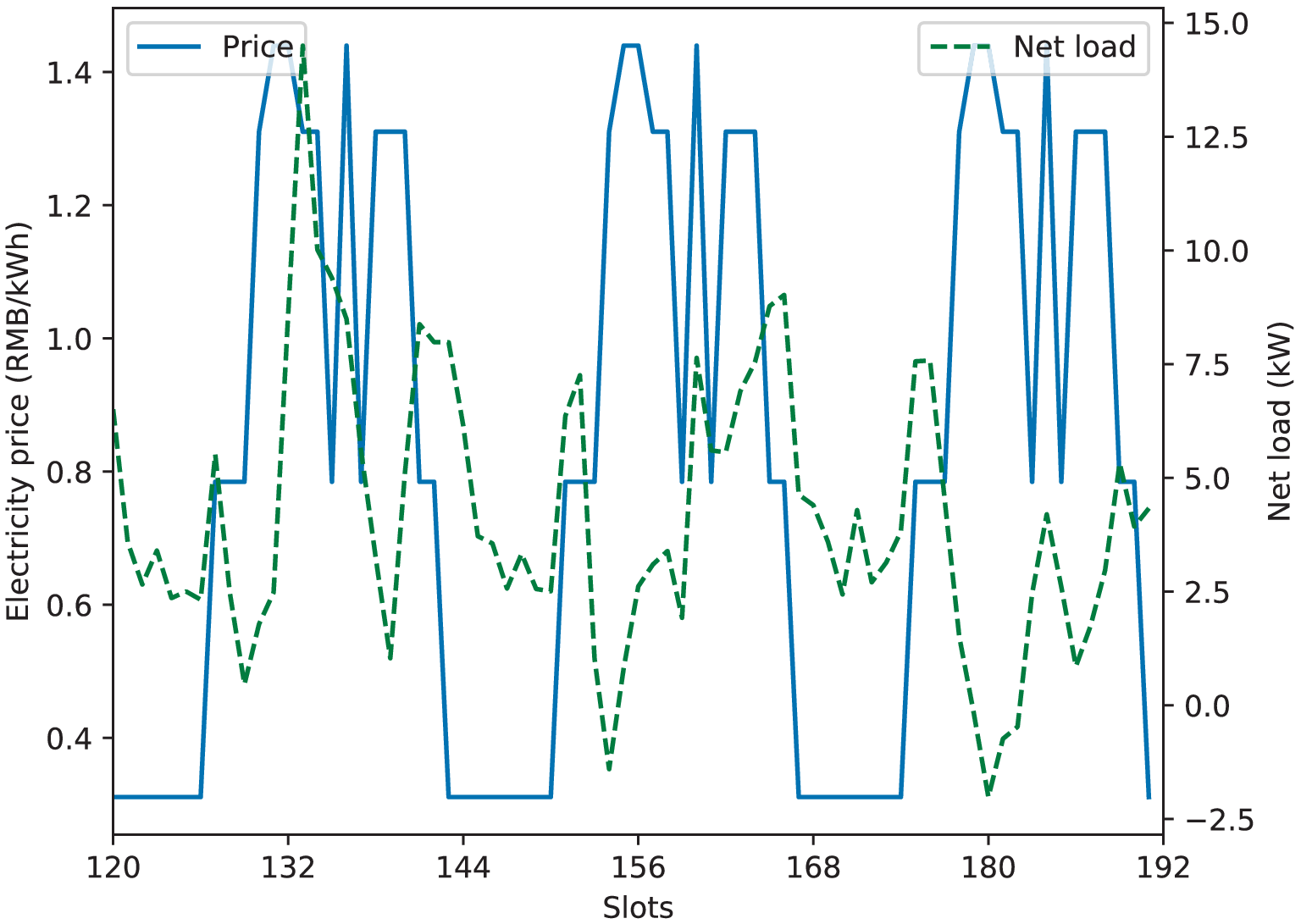}
\end{minipage}}
\subfigure[BESS action]{
\begin{minipage}[b]{0.23\textwidth}
\includegraphics[width=1\textwidth]{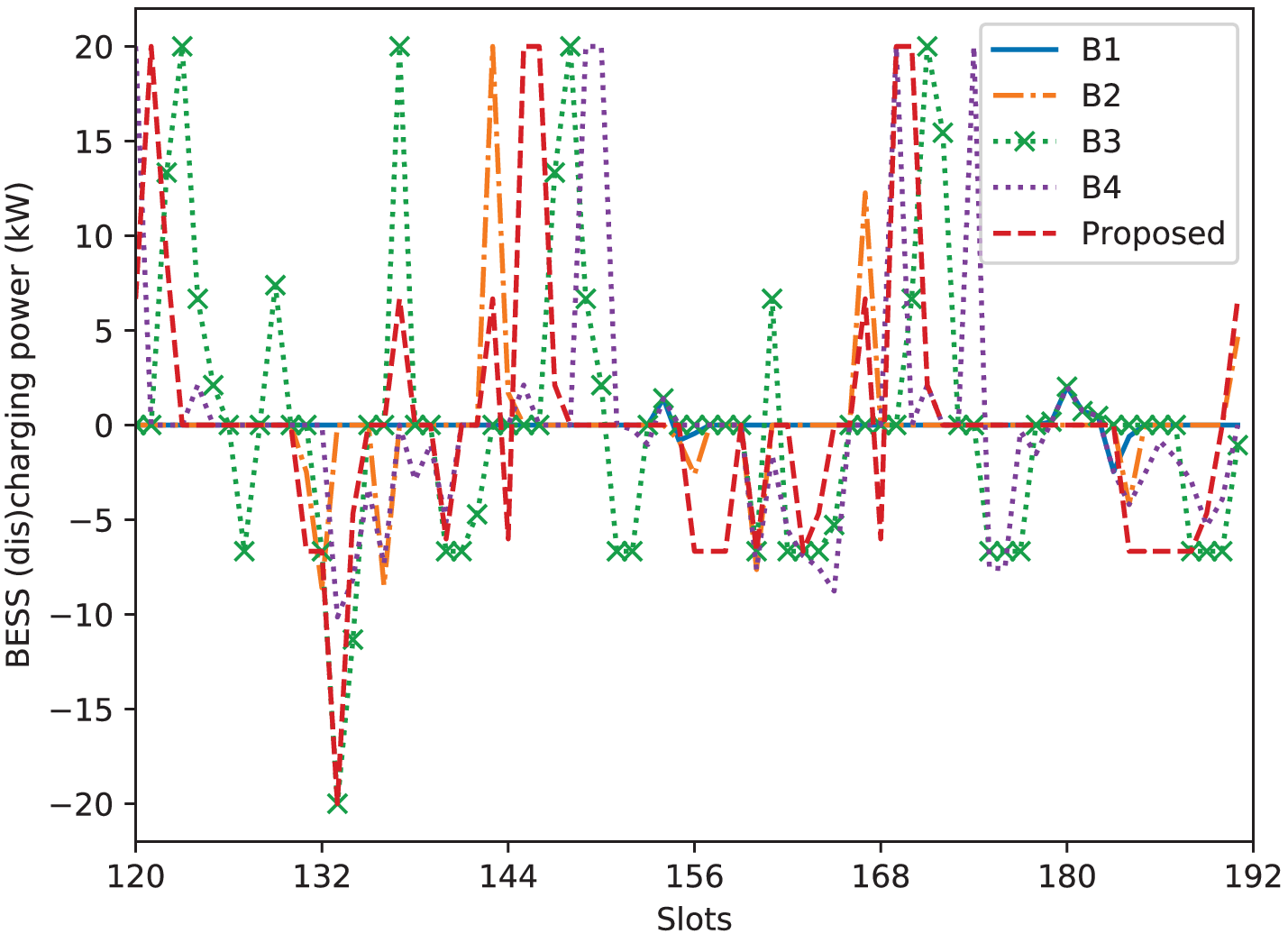}
\end{minipage}}
\subfigure[HESS action]{
\begin{minipage}[b]{0.23\textwidth}
\includegraphics[width=1\textwidth]{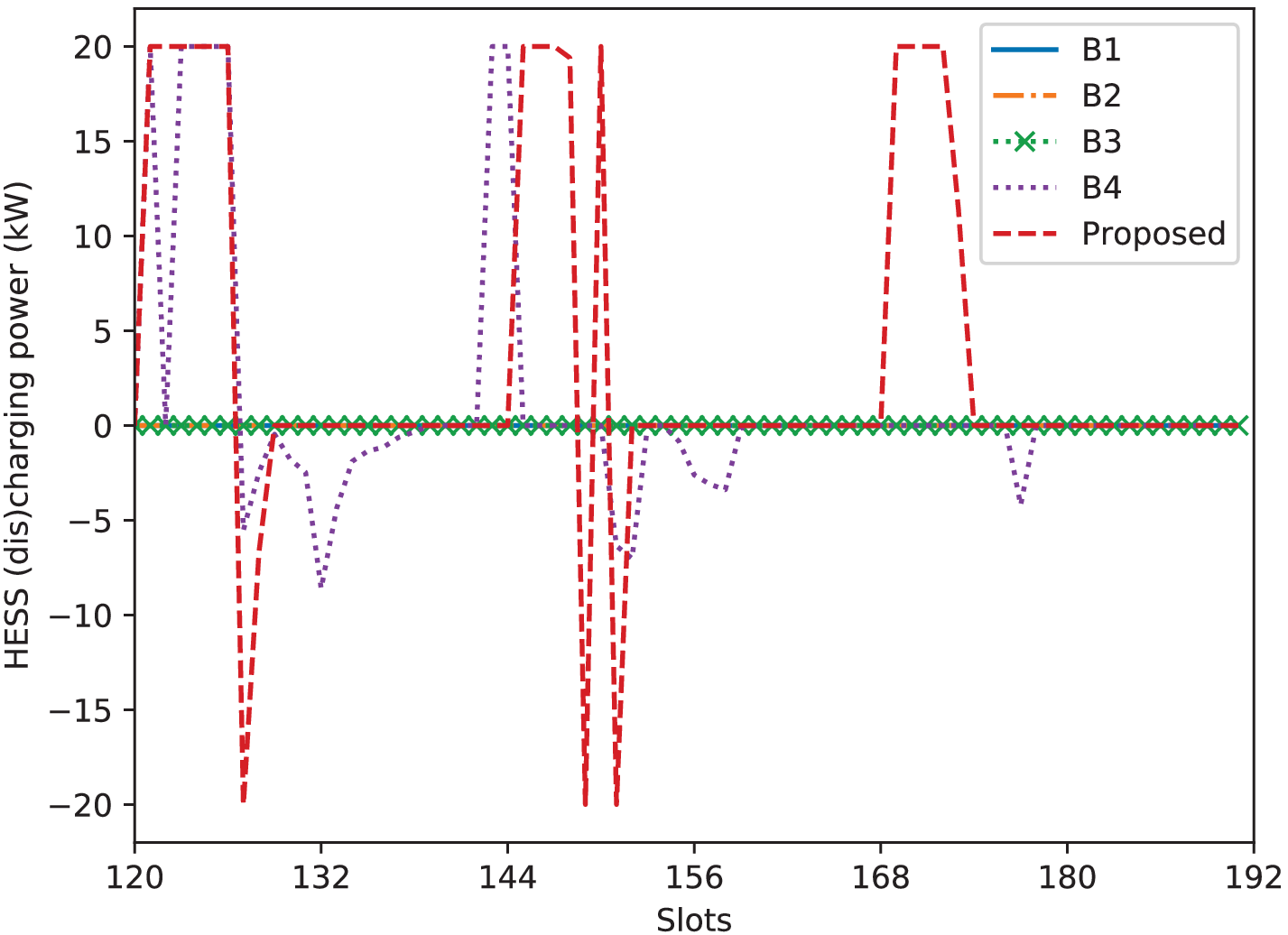}
\end{minipage}}
\subfigure[Temperature]{
\begin{minipage}[b]{0.23\textwidth}
\includegraphics[width=1\textwidth]{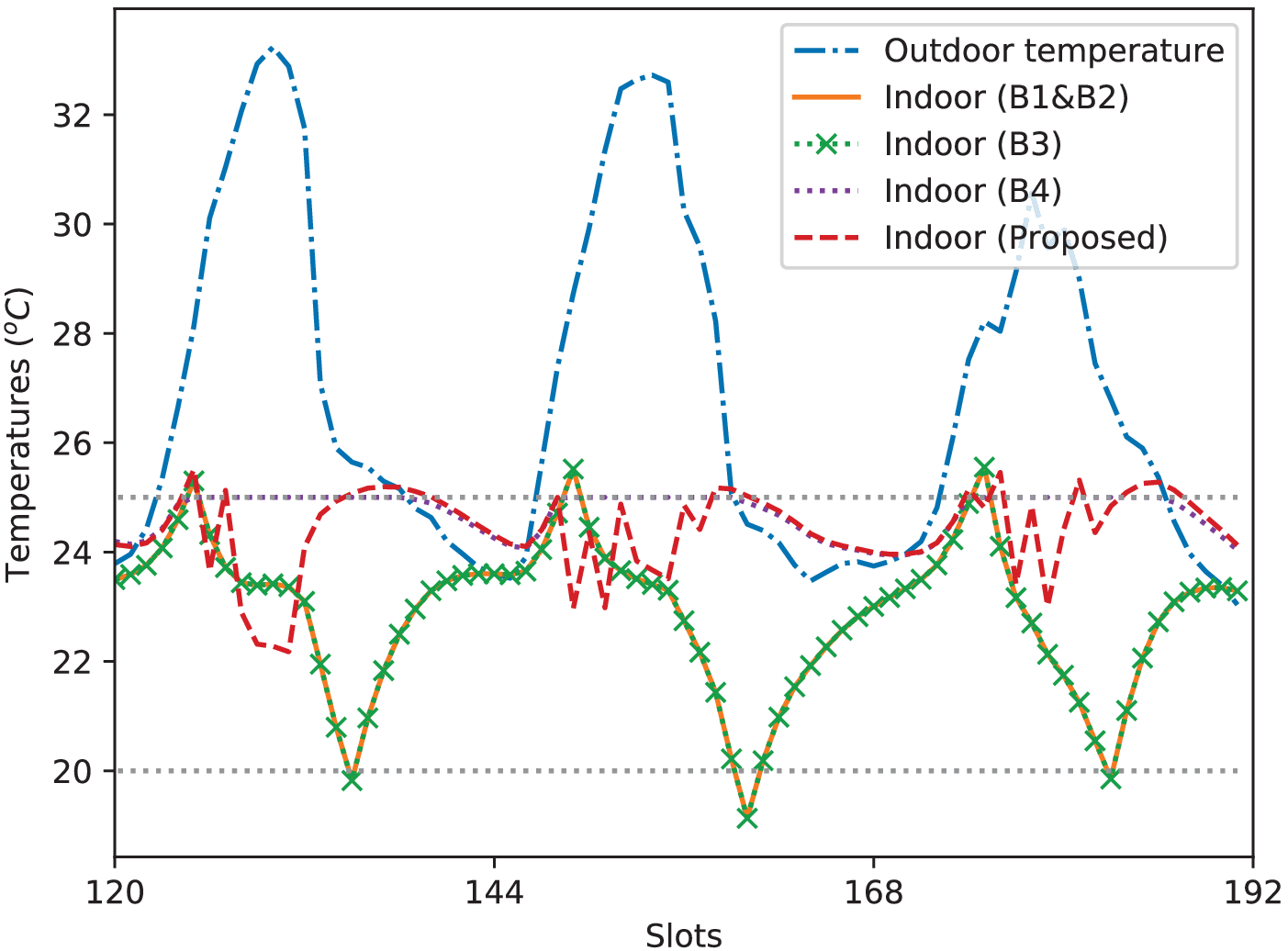}
\end{minipage}}
\caption{Performance details under case-1.} \label{fig_6}
\end{figure*}

\begin{figure*}
\centering
\subfigure[Price and net load]{
\begin{minipage}[b]{0.23\textwidth}
\includegraphics[width=1\textwidth]{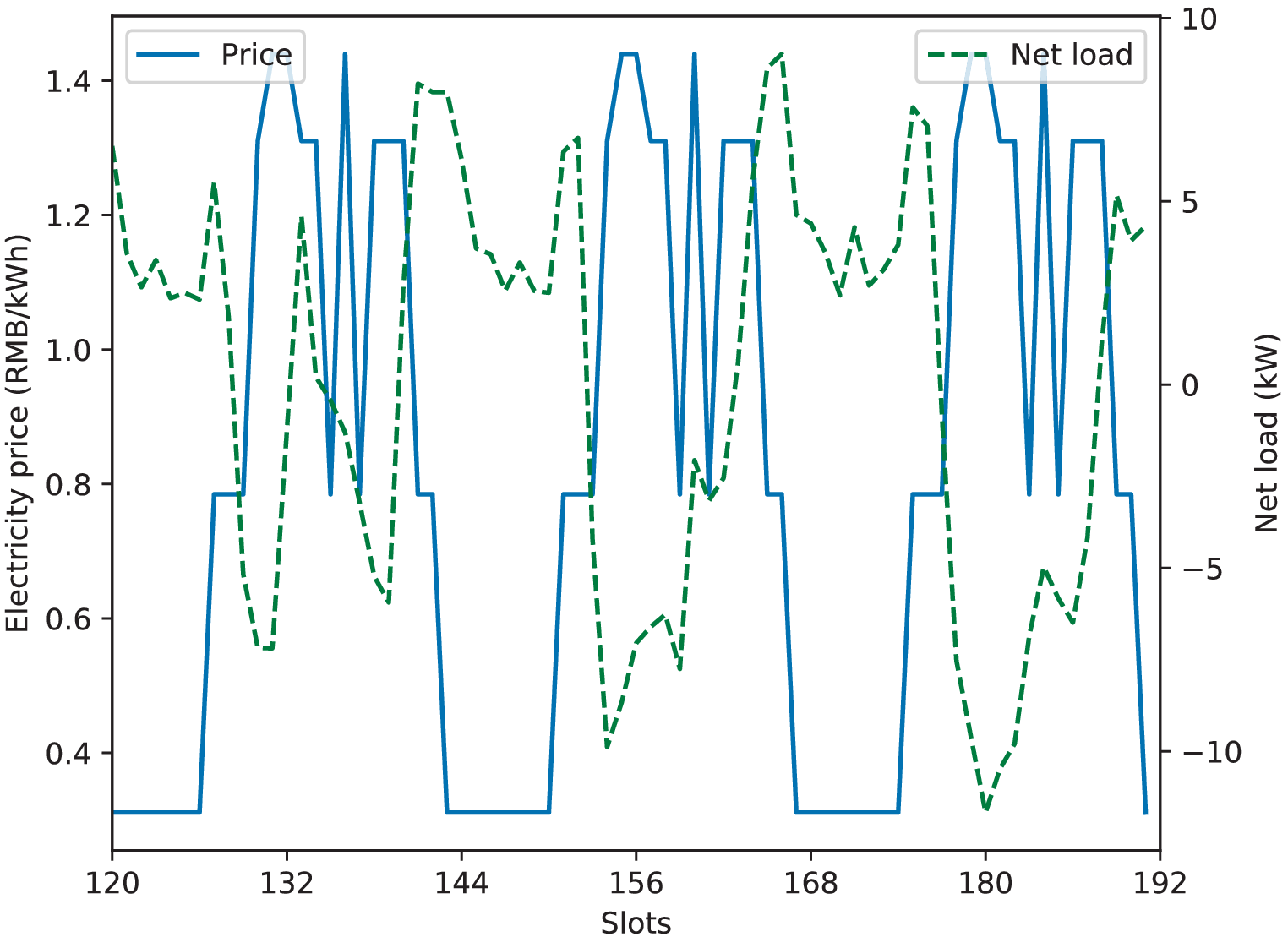}
\end{minipage}}
\subfigure[BESS action]{
\begin{minipage}[b]{0.23\textwidth}
\includegraphics[width=1\textwidth]{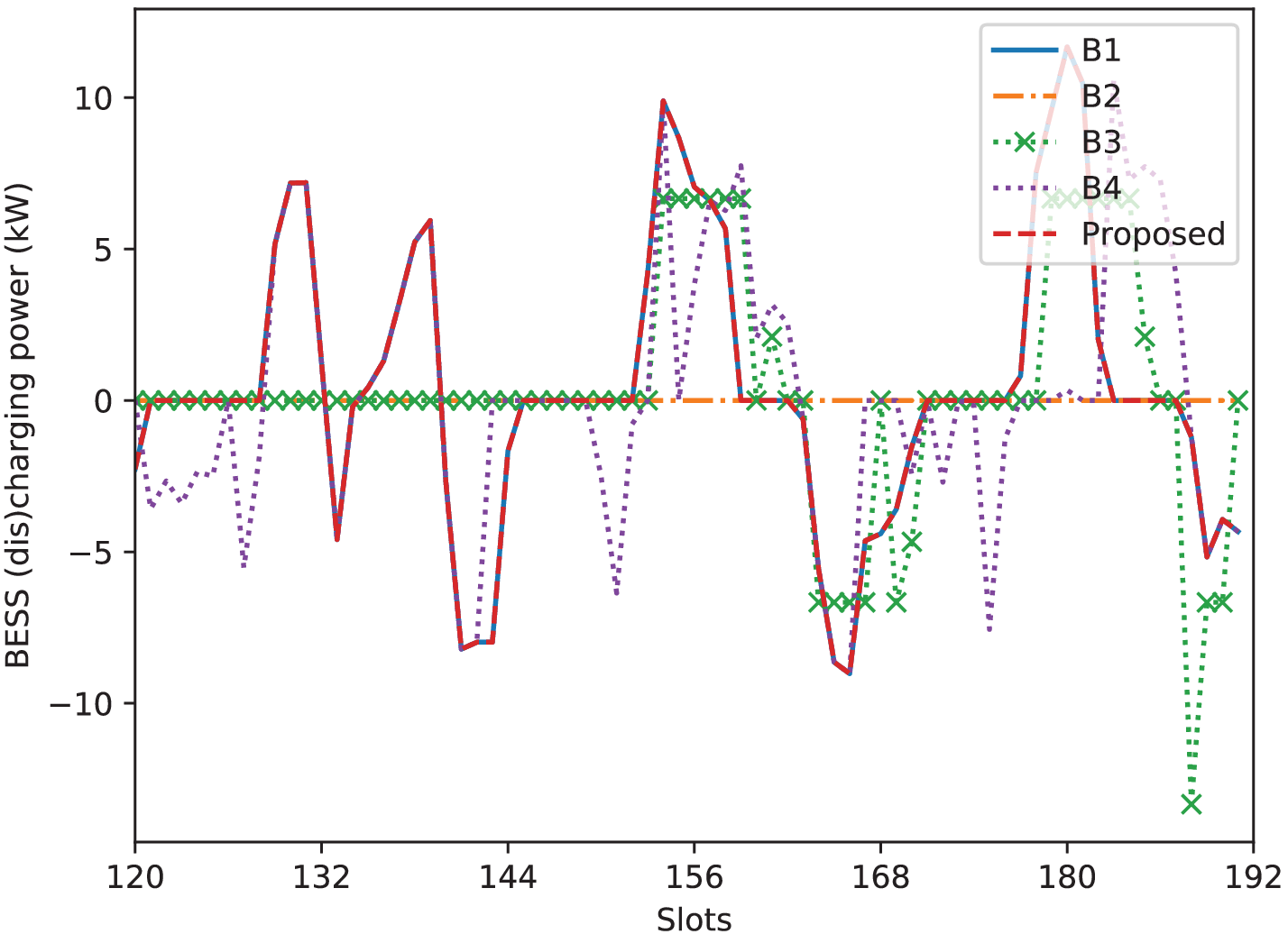}
\end{minipage}}
\subfigure[HESS action]{
\begin{minipage}[b]{0.23\textwidth}
\includegraphics[width=1\textwidth]{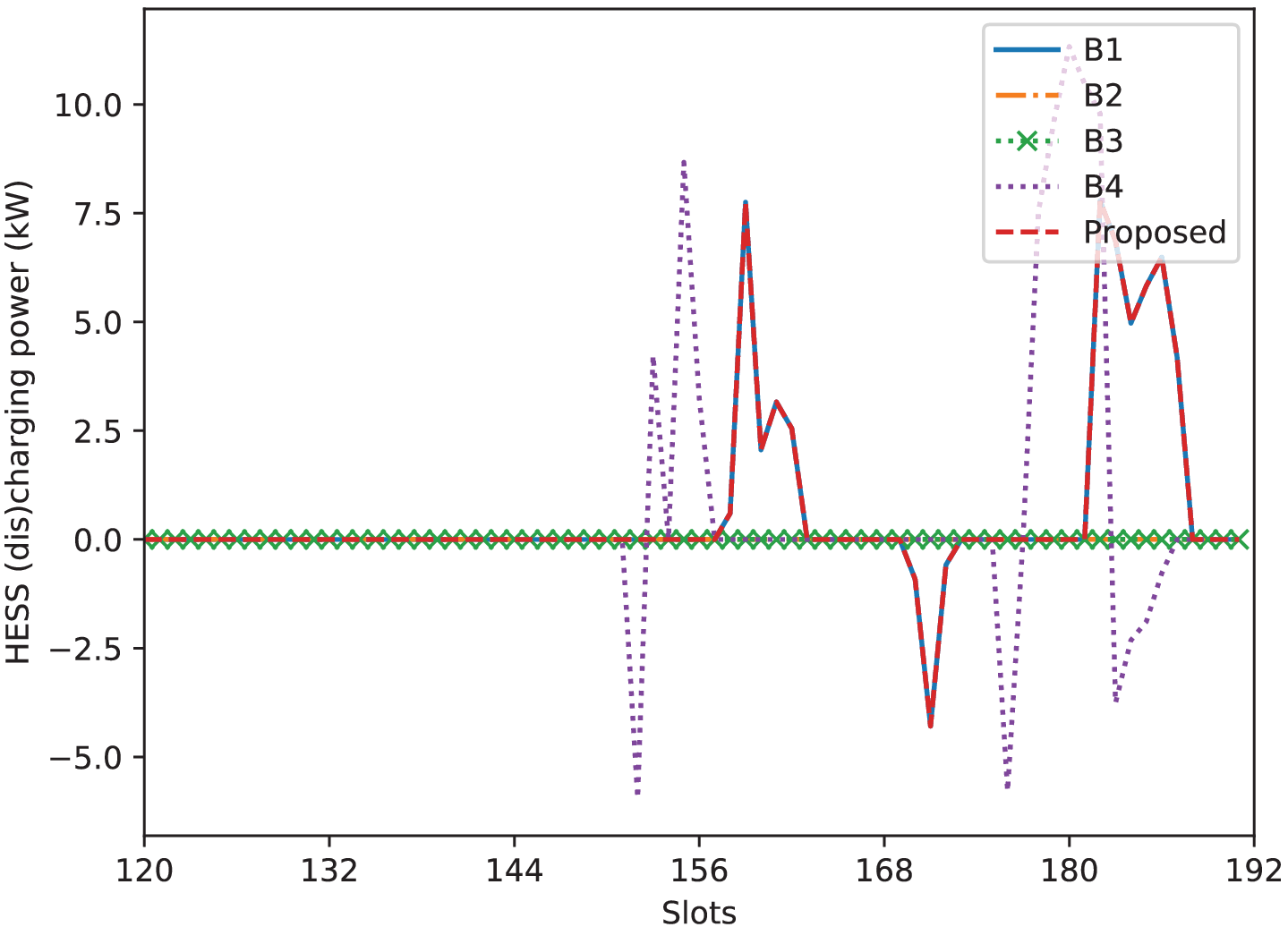}
\end{minipage}}
\subfigure[Temperature]{
\begin{minipage}[b]{0.23\textwidth}
\includegraphics[width=1\textwidth]{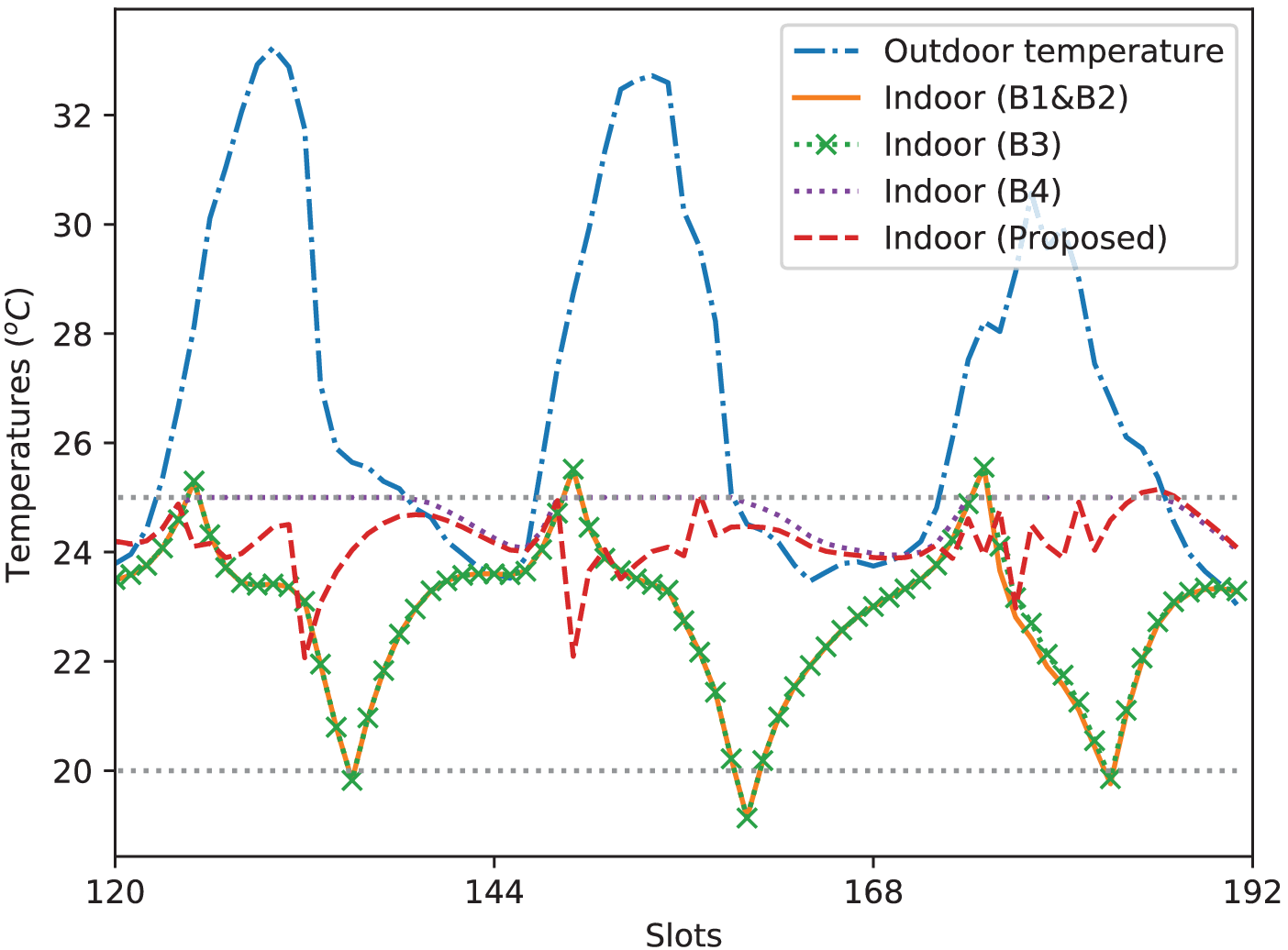}
\end{minipage}}
\caption{Performance details under case-2.} \label{fig_7}
\end{figure*}

In Figs.~\ref{fig_6} and \ref{fig_7}, more performance details are presented. Since price information is not utilized, \textbf{B1} may buy electricity at high prices when net load is greater than zero under case-1. As a result, \textbf{B1} has the highest electricity cost as shown in Fig.~\ref{fig_5}(c). In contrast, \textbf{B2} can avoid the above action by charging BESS at the minimum price and discharging BESS at the maximum price. Compared with \textbf{B2}, electricity cost under \textbf{B3} can be reduced by scheduling BESS and HESS jointly using DDQN, which can find more opportunities of discharging/charging electricity for cost reduction as shown in Figs.~\ref{fig_6}(b) and (c). Compared with \textbf{B3}, thermal loads and BESS/HESS are scheduled jointly under the proposed algorithm. As depicted by Fig.~\ref{fig_6}, HESS generates hydrogen when prices are low and consumes hydrogen for heat and electricity generation when BESS is idle at relatively high prices and/or thermal inputs are required by buildings. Moreover, indoor temperatures under the proposed algorithm are close to $\beta_i^{\max}$ similar to \textbf{B4}, resulting in reduced cost of natural gas as shown in Fig.~\ref{fig_6}(d) and Fig.~\ref{fig_5}(c). Since perfect information is assumed by \textbf{B4}, indoor temperatures under \textbf{B4} can be adjusted to $\beta_i^{\max}$ accurately when outdoor temperature is higher than $\beta_i^{\max}$ as depicted by Fig.~\ref{fig_6}(d).

Different from case-1, there are many surplus renewable energies at high prices under case-2 as shown in Fig.~\ref{fig_7}(a). Therefore, \textbf{B2} and \textbf{B3} have few opportunities of reducing electricity cost by discharging BESS and HESS at high prices. In contrast, \textbf{B1} can store excess energy in BESS and HESS for future use, e.g., discharging BESS and HESS when net load is greater than zero. As a result, \textbf{B1} has lower operational cost than \textbf{B2} and \textbf{B3} as shown in Fig.~\ref{fig_5}(a). Since the proposed algorithm uses the same rule as \textbf{B1} to schedule BESS/HESS and schedules thermal loads based on multi-agent discrete actor-critic, the proposed algorithm achieves lower operational cost than \textbf{B1} as shown in Fig.~\ref{fig_5}(a).

\subsection{The impact of thermal disturbances}\label{s55}

\begin{table*}[htb]
\renewcommand{\arraystretch}{1.3}
\caption{Performances under varying thermal disturbances} \label{table_2} \centering
\begin{tabular}{|c|c|c|c|c|c|c|c|c|c|c|c|c|}
\hline  & \multicolumn{6}{c|}{Case-1} & \multicolumn{6}{c|}{Case-2}\\
\hline  & \multicolumn{3}{c|}{Operational cost (RMB)} &  \multicolumn{3}{c|}{ATD ($^oC$)}  & \multicolumn{3}{c|}{Operational cost (RMB)} &  \multicolumn{3}{c|}{ATD ($^oC$)} \\
\hline  \textbf{B1}    & 3320.000 & 3326.000 & 3477.000 & 0.132 &0.282 & 0.571    & 1587.000 &1593.000 &1731.000 &0.124 &0.286 &0.521 \\
\hline  \textbf{B2}    & 3285.000 & 3291.000 & 3441.000 & 0.132 &0.282 & 0.571    & 2200.000 &2206.000 &2347.000 &0.132 &0.282 &0.521\\
\hline  \textbf{B3}    & 2897.000 & 2968.000 & 3192.000 & 0.149 &0.302 & 0.593    & 2041.000 &1814.000 &1986.000 &0.133 &0.284 &0.571\\
\hline  \textbf{B4}    & - & - & - & - & - & -                                    & -  & - & - & - & - & -  \\
\hline  \textbf{Proposed}  & 2757.000 & 2822.000 & 2683.000 & 0.117 &0.328 & 0.652    & 1166.000 &1180.000 &1305.000 &0.140 &0.260 &0.630\\
\hline
\end{tabular}
\end{table*}

The performances of all schemes under varying uncertain thermal disturbance are shown in Table~\ref{table_2}, where thermal disturbance $\varrho_{t}$ is assumed to follow a uniform distribution with parameters -$\chi$ and $\chi$ and three scenarios are considered, i.e., $\chi$$\in$$\{0.9,1.8,2.4\}^o$F. It can be seen that the proposed algorithm can achieve lower operational cost than \textbf{B1}-\textbf{B3} with a small sacrifice even no sacrifice of ATD. Moreover, \textbf{B4} has no feasible solution when any violation of comfortable temperature range is not allowed. In summary, the proposed algorithm is still useful for operational cost reduction even if there are thermal disturbances in the environment.

\section{Conclusions and Future Work}\label{s6}
In this paper, we investigated an optimal operation problem of an HBMES and proposed an energy management algorithm to solve the problem based on multi-agent discrete actor-critic with rules. The proposed algorithm does not require any prior knowledge of uncertain parameters, parameter prediction, and explicit building thermal dynamics models. Simulation results showed the effectiveness of the proposed algorithm. In future work, we intend to investigate the optimal operation problem of the HBMES with more kinds of factors, e.g., direct hydrogen selling/buying, heat demands, and controllable electric loads. Since the adoption of distributed energy sources (e.g., ESSs) will incur investment cost, it is important to study the optimal planning problem of an HBMES so that the expected system expenditure can be minimized.

\end{document}